\documentclass
 [aps,prl,reprint,preprintnumbers,showpacs,floatfix,nofootinbib,superscript
 address,longbibliography]{revtex4-2}
\usepackage[utf8]{inputenc}
\usepackage[T1]{fontenc}
\usepackage{amsmath,amsfonts,amssymb}
\usepackage{epsf,amsmath,bbold,amsfonts,stmaryrd}
\usepackage{hhline}
\usepackage{mathrsfs}
\usepackage{mathtools}
\usepackage[dvipsnames]{xcolor}
\usepackage{multirow,tabularx}
\usepackage{graphicx}
\usepackage{natbib}

\usepackage{xspace}
\usepackage{xstring}
\usepackage{titlesec}
\usepackage{parskip}
\usepackage{calc}
\usepackage{bm}
\usepackage{float}

\allowdisplaybreaks
\usepackage{braket}

\newrobustcmd{\pea}[1]{%
 	\emph{#1}\textbf{\ \ \ ---}
 }
\titleformat{\paragraph}[runin]{\normalfont\normalsize\bfseries}
 {\emph\theparagraph}{1em}{\pea}

\newcommand*{\ie}{i.e.\@\xspace}
\newcommand*{\eg}{e.g.\@\xspace}

\newcommand*{\eq}{eq.\@\xspace}

\def\a{\alpha}

\def\f{\frac}

\def\mc{\mathcal}

\def\m{\mu}
\def\n{\nu}

\def\p{\partial}

\def\s{\sigma}

\def\x{\xi}

\def\be{\begin{equation}}
\def\ee{\end{equation}}

\def\bea{\begin{eqnarray}}
\def\eea{\end{eqnarray}}

\def\ba{\begin{array}}
\def\ea{\end{array}}

\def\bc{\begin{center}}
\def\ec{\end{center}}

\def\bl{\begin{flushleft}}
\def\el{\end{flushleft}}

\def\br{\begin{flushright}}
\def\er{\end{flushright}}

\def\bi{\begin{itemize}}
\def\ei{\end{itemize}}

\def\bt{\begin{tabular}}
\def\et{\end{tabular}}

\def\be{\begin{equation}}
\def\ee{\end{equation}}

\def\bea{\begin{eqnarray}}
\def\eea{\end{eqnarray}}

\def\f{\frac}

\def\p{\partial}

\usepackage{hyperref}
\hypersetup{%
     pageanchor= false,%
     colorlinks = true,%
     allcolors= red}%

\usepackage{xifthen}
\newcommand*\diff{\mathrm{d}} 
\newcommand*\ldiff[2][]{ \ifthenelse{\isempty{#1}}{ \diff
#2}{\diff^#1#2} \,} 
\let\limitint\int 
\renewcommand{\int}{\limitint \!} 

\makeatletter 
    
\renewcommand\onecolumngrid{%
\do@columngrid{one}{\@ne}%
\def\set@footnotewidth{\onecolumngrid}%
\def\footnoterule{\kern-6pt\hrule width 1.5in\kern6pt}%
}%

\makeatother    

\begin{document}

\title{A non-compact QCD axion}

\author{Georgios K. Karananas}
\email{georgios.karananas@physik.uni-muenchen.de}
\affiliation{Max-Planck-Institut für Physik, Boltzmannstraße 8, 85748 Garching
 bei M\"unchen, Germany}
\affiliation{Arnold Sommerfeld Center, Ludwig-Maximilians-Universit\"at,
 Theresienstraße 37, 80333 M\"unchen, Germany}
\author{Mikhail Shaposhnikov}
\email{mikhail.shaposhnikov@epfl.ch}
\affiliation{Institute of Physics, \'Ecole Polytechnique F\'ed\'erale de
 Lausanne, CH-1015 Lausanne, Switzerland}
\author{Sebastian Zell}
\email{sebastian.zell@lmu.de}
\affiliation{Arnold Sommerfeld Center, Ludwig-Maximilians-Universit\"at,
 Theresienstraße 37, 80333 M\"unchen, Germany}
\affiliation{Max-Planck-Institut für Physik, Boltzmannstraße 8, 85748 Garching
 bei M\"unchen, Germany}

\begin{abstract}
We investigate the cosmology of an axion that is fundamentally non-compact.
During inflation, fluctuations of the effectively massless field populate
many QCD vacua, thereby evading conventional isocurvature constraints while
generating domain walls---without accompanying cosmic strings. A small
non-QCD contribution to the axion potential is required to trigger the timely
collapse of domain walls; as a consequence, a residual amount of CP violation
in the strong sector must exist, potentially within reach of planned
experiments. Non-compact axions can account for the entirety of the dark
matter abundance, and the collapse of domain walls sources a stochastic
gravitational-wave background at nanohertz frequencies. Such axion dynamics
can be embedded in top-down constructions---such as Weyl-invariant
Einstein–Cartan gravity---where the tilting of the axion potential arises
automatically.
\end{abstract}
\maketitle

\paragraph*{Introduction and summary}

The QCD axion~\cite{Peccei:1977hh,Weinberg:1977ma,Wilczek:1977pj} remains one
of the most compelling candidates for physics beyond the Standard Model.
Originally proposed as a dynamical solution to the strong CP problem, it also
provides a viable explanation for the dark matter (DM) observed in our
Universe. In its canonical realization, the axion emerges as the
pseudo-Nambu-Goldstone boson of a spontaneously broken Peccei-Quinn (PQ)
symmetry~\cite{Peccei:1977hh}. Beyond this framework, other mechanisms to
generate an axion have been proposed, including realizations based on extra
dimensions~\cite{Witten:1984dg} (see the review~\cite{Reece:2024wrn}) and on
local gauge
invariance~\cite{Dvali:2005an,Dvali:2013cpa,Dvali:2017mpy,Dvali:2022fdv}.

A common feature of most scenarios is that the axion is a compact field, with
a periodic field space. In contrast, in this work we explore the possibility
that the axion is fundamentally non-compact. Such a setup receives
motivation, for instance, from the possibility that the axion originates from
gravitational
dynamics~\cite{Mercuri:2009zi,Karananas:2024xja,Karananas:2025ews}. In a
non-compact field space, contributions to the axion potential are not
restricted to those generated by QCD alone. As we show, this qualitatively
alters the cosmological history of the axion and leads to a set of
experimental signatures distinct from those of the standard variants of the
QCD axion (see
\eg~\cite{Marsh:2015xka,DiLuzio:2020wdo,Choi:2020rgn}). In this paper, we
present the cosmology of a non-compact axion using an effective field theory
description, without committing to a specific ultraviolet completion.

During inflation, the axion is taken to be effectively massless, so that
quantum fluctuations populate its field space on superhorizon scales. At
later times, nonperturbative QCD effects generate the familiar periodic axion
potential, giving rise to a discretuum of vacua separated by $2\pi f_a$,
where $f_a$ denotes the low-energy decay constant. We show that, despite the
non-compact nature of the field, $f_a$ can be chosen such that the axion
accounts for the entire DM abundance.

A  massless axion during inflation generically induces isocurvature
perturbations~\cite{Turner:1990uz}, which are tightly constrained by
observations of the cosmic microwave background (CMB)~\cite{Planck:2018jri}.
We demonstrate that these bounds can be evaded in our framework. The key
point is that inflationary fluctuations populate a large number of distinct
QCD vacua~\cite{Kofman:1985zx,Kofman:1986wm,Linde:1990yj}, a direct
consequence of the non-compact field space. This process inevitably leads to
the formation of topological defects. In the absence of an underlying U(1)
symmetry, cosmic strings do not form, and the only defects present are domain
walls.

Stable domain walls are cosmologically unacceptable, as they would rapidly
dominate the energy density of the Universe~\cite{Sikivie:1982qv}.
Consistency therefore requires a small non-QCD contribution to the axion
potential that lifts the degeneracy between neighboring vacua and triggers
the collapse of the domain-wall network. We show that this can naturally
occur before Big Bang Nucleosynthesis (BBN), ensuring the viability of the
scenario. The annihilation of domain walls sources a stochastic background of
gravitational waves, with a present-day peak frequency in the nanohertz range
and an energy density of order $10^{-10}$. While this signal lies below
current sensitivities, it may become accessible to future pulsar timing
experiments.

The same mechanism responsible for lifting the vacuum degeneracy also induces
CP violation in the strong sector. We find that the resulting effective
$\theta$ angle is at most two orders of magnitude below current
bounds~\cite{Abel:2020pzs} from the neutron electric dipole moment, assuming
the axion constitutes all of DM. Upcoming proton electric dipole moment
experiments~\cite{Anastassopoulos:2015ura}, which aim to improve sensitivity
by roughly three orders of magnitude, will therefore directly probe this
scenario. An observation of strong CP violation at this level would
constitute a striking signature of non-compact axion dynamics.

Finally, we note that the required axion dynamics arises naturally in
theories where the axion has a gravitational origin. As an explicit example,
we discuss an embedding based on Weyl-invariant Einstein–Cartan
gravity~\cite{Karananas:2024xja,Karananas:2025ews}. In this context, the
presence of a non-QCD contribution to the axion potential is unavoidable,
while its smallness—mandated by experimental constraints—is directly tied to
the weakness of gravitational couplings. From this perspective, the scenario
presented here provides a coherent link between axion dark matter, primordial
gravitational waves, and CP violation in QCD.

\paragraph*{The effective theory}

We consider a noncompact dimensionless field $\theta$ with effective
Lagrangian\,\footnote{We take the metric to be mostly plus.}
\be
\label{eq:ALP_generic_Lagrangian}
\mathcal L = - \f{f^2(\Phi)}{2}\p_\m\theta \p^\m \theta 
+\f{\theta}{F(\Phi)}Q - \delta V  \ ,
\ee
where
\be
Q = \f{\alpha_s}{8 \pi}G^b_{\m\n}\widetilde G^{b\,\m\n} \ , 
~~~\a_s = \f{g_s^2}{4\pi} \ .
\ee
Here $G_{\m\n}$ is the field strength tensor of QCD, $\widetilde G_{\m\n} =
1/2\, \epsilon_{\m\n\rho\s}G^{\rho\s}$ denotes its dual, $g_s$ is the strong
coupling, and summation over color index $b$ is understood. Furthermore,
$\delta V$ represents a non-QCD related contribution to the potential, and
for concreteness we consider a mass term
\be
\label{eq:potential_tilt}
\delta V = \f{f_m^2(\Phi)\widetilde m_a^2}{2}\theta^2 \ ,
\ee 
as also inspired by the situation in Weyl-invariant Einstein-Cartan
gravity~\cite{Karananas:2024xja,Karananas:2025ews}. Generalizing to different
types of non-QCD related contributions is straightforward.

The above theory is specified by three functions $f,~f_m$ and $F$; the former
two carry mass-dimension one and the latter is dimensionless. In general,
$f,~f_m$, and $F$ depend on the various fields present---such as $\theta$
itself and the inflaton---denoted collectively by $\Phi$. We require that
$f,~f_m$, and $F$ take constant values after inflation and reheating. If
these function change rapidly, \eg during reheating, this can lead to an
overproduction of axions and potentially a violation of the effective field
theory.\footnote{For $f_I\gg f_{\rm QCD}$, an overproduction has been
observed~\cite{Kawasaki:2013iha,Harigaya:2015hha,Kearney:2016vqw,Kobayashi:2016qld,Co:2017mop,Co:2020dya,Ballesteros:2021bee,Graham:2025iwx}.
In non-minimally coupled inflationary setups, however, one generically has
$f_I < f_{\rm QCD}$~\cite{Rigouzzo:2025hza, Rigouzzo:2025ycb} (as well
as~$f_{m,I} < f_{m,{\rm QCD}}$); we checked, following
\cite{Garcia-Bellido:2012npk}, that no overproduction of axions occurs in
this case.}

Since our proposed mechanism is fully operative also if $f$, $f_m$, and $F$
do not vary, we shall assume in the following that these be constant or only
slowly varying. To keep the discussions general, we shall keep track of the
epoch at which the different functions are evaluated. In a self-explanatory
notation, the subscripts ${\rm QCD}$ and $I$ denote quantities evaluated at
late and inflationary times, respectively. Without loss of generality we take
$F(\Phi_{\rm QCD})\equiv F_{\rm QCD}=1$, which we can achieve by rescaling
$\theta \rightarrow F_{\rm QCD} \theta$, and redefining $f_{\rm QCD}$, $f_I$
and $F_I$.

\paragraph*{CP-violation} 

As an immediate concern, even a tiny (in this case
``gravitationally-induced'') non-QCD mass for the axion, necessarily
translates into observable CP-violation: the ground state is shifted away
from the parity-preserving point. This becomes obvious by inspecting the
potential, which at small field values is well approximated by
\be \label{biasedPotential}
V_{\rm QCD}(\theta) +\delta V  \approx 
\f{f_a^2 M_a^2}{2} \left((\theta-\bar \theta)^2 + \m^2 \theta^2\right) \ ,
\ee
where we introduced
\be
\label{eq:mass_ratio}
\m = \f{m_a}{M_a} \ ,~~~
m_a = \widetilde m_a \f{f_{m,{\rm QCD}}}{f_a} \ ,~~~ 
f_a = f_{\rm QCD} \ ,
\ee
$f_a$ acts as effective late-time decay constant, and $M_a$ is the
QCD-induced axion mass~\cite{ParticleDataGroup:2024cfk}
\be
\label{eq:QCD_mass}
M_a \simeq 
5.7\times10^{-15}~ {\rm GeV} \left(\f{10^{12}~ {\rm GeV}}{f_a}\right) \ .
\ee

In \eq~(\ref{biasedPotential}), $|\bar \theta| \sim \mc O(1)$, as it
comprises the QCD vacuum angle and the quark mass phase. The physical
CP-violating parameter is
\be \label{CPViolation}
\theta_{\rm phys} \propto \m^2 +\ldots \ ,
\ee
where the ellipses are higher order in $\m^2$. The experimental bound on the
neutron electric dipole moment~\cite{Abel:2020pzs},
\be
\label{eq:EDM_bound}
\theta_{\rm phys}^{\rm exp} \lesssim 10^{-10} \ ,
\ee
dictates that
\be
\label{mubound}
\mu^2 \lesssim 10^{-10} \,.
\ee

\paragraph*{Inflation}

Turning to the early Universe, we will assume that $\theta$ is practically
massless during inflation and we envisage a high-scale inflationary epoch
(\eg driven by the Higgs field~\cite{Bezrukov:2007ep}), with $H_I$ the
corresponding Hubble parameter. The amplitude of primordial fluctuations
$\delta \theta_I$ of the massless spectator $\theta$ is of order
\be
\label{eq:amplitude_theta_I}
\delta\theta_I \simeq \f{H_I}{2\pi f_I} \ .
\ee
Eventually, the perturbations $\delta\theta_I$ freeze-out and assume a
constant value.

The nonrenormalizable coupling of the axion to the QCD topological invariant
implies that the effective field theory is valid up to a certain cutoff
$\Lambda_I$. The dominant process that determines this scale is gluon-gluon
scattering via $\theta$ exchange. A computation of the corresponding
amplitude (see appendix~\ref{app:effectiveTheory}) shows that perturbative
unitarity is saturated at energies in the vicinity of
\be
\Lambda_I \simeq \f{8\pi^{3/2}}{\a_s} f_I F_I \ .
\ee
Importantly, the non-compactness of $\theta$ increases the cutoff
scale~\cite{Zell:2024cyz}: $\Lambda_I\sim f_I F_I/\alpha_s$, since this is
the scale suppressing the non-renormalizable operator in
\eq~(\ref{eq:ALP_generic_Lagrangian}). In contrast, a compact PQ axion would
have a lower $\Lambda_I\sim f_I F_I$, where the radial mode becomes active.

As a benchmark, we take the inflationary scale $\sim \mc O(10^{13})~{\rm
GeV}$, meaning that $\a_s \sim 1/40$ (see
\eg~\cite{Berghaus:2025dqi}), and the above gives
\be
\Lambda_I \simeq \mc O({\rm few}) \times 10^3 f_I F_I \ .
\ee
Therefore, a necessary requirement for self-consistency is that $\Lambda_I$
be larger than the inflationary Hubble scale, yielding
\be
\label{eq:EFT_constraint}
f_I F_I > 10^{-3} H_I \ .
\ee
In general, reheating leads to a stronger constraint since the reheating
temperature is larger, $T_{\rm reh}>H_I$ (and $\Lambda$ can differ from
$\Lambda_I$). So the dynamics of reheating, which depend on the specific
inflationary model, have to be determined in order to check the validity of
the effective field theory.

\paragraph*{Isocurvature perturbations}

\begin{figure}
\centering 
\includegraphics[scale=.25]{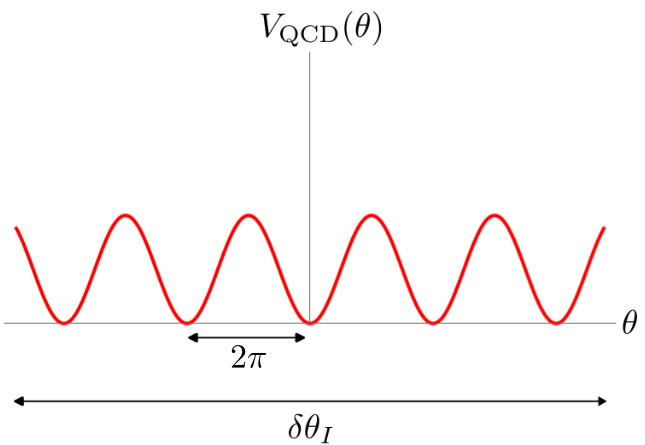}
\caption{The inflationary fluctuations $\delta\theta_I$ are taken to span a
plethora of vacua.}
\label{fig:axion_potential_standard}
\end{figure}

At times $t_{\rm QCD}\gg t_I$, the usual axionic potential $V_{\rm
QCD}(\theta)$ is generated. Around its minimum, it is approximated by
\eq~(\ref{biasedPotential}). Taking into account its full form, we know that
the distance between minima is $\delta\theta_{\rm QCD} = 2\pi$, as
illustrated in Fig.~\ref{fig:axion_potential_standard}.

Now we consider the case that the modes $\delta \theta_I$ created during
inflation, see~(\ref{eq:amplitude_theta_I}), are larger than
$\delta\theta_{\rm QCD}$. Then isocurvature bounds from CMB are completely
evaded: the primordial axion perturbations feel many vacua, meaning that they
imprint on the spectrum as white noise, and their contribution is therefore
suppressed~\cite{Kofman:1985zx,Kofman:1986wm,Linde:1990yj}. Indeed,
generalizing the considerations of~\cite{Linde:1990yj} to our setup, the
isocurvature density perturbations on a comoving scale $l$ generated by
misalignment, as long as $H_I>f_I$, are given by
\be
\label{eq:isocurv_pert_CMB}
\f{\delta\rho}{\rho}\Bigg\vert_{\rm mis.} 
\sim \f{H_I}{2\pi  f_I}
\left(\f{l}{l_c}\right)^{-\f{H_I^2}{8\pi^2 f_I^2 }} 
\Omega_\theta \ ,
\ee
where $l_c\sim 10^{17}~{\rm cm}$ is the present-day size of the modes created
around the QCD scale, and $\Omega_\theta$ is the axion abundance. For CMB
modes, $l_{\rm CMB}
\sim 10^{28}~{\rm cm}\gg l_c$, so from~(\ref{eq:isocurv_pert_CMB}) we find
that the isocurvature perturbations are suppressed provided
that\,\footnote{Note that a similar to~(\ref{eq:Hubble_constraint_1})
restriction can be found by taking at face value that $\delta \theta_I >
\delta\theta_{\rm QCD}$, translating into $H_I/ f_I > (2\pi)^2$.}
\be
\label{eq:Hubble_constraint_1}
H_I  \gtrsim 2\sqrt{2}\pi f_I \ . 
\ee
Even for constant $F_I=1$, this is compatible with the effective field theory
constraint~(\ref{eq:EFT_constraint}). The fact that the cutoff scale is
higher for the non-compact axion as compared to the PQ case, $\Lambda_I \sim
f_I F_I/\alpha_s$ instead of $\Lambda_I \sim f_I F_I$, is crucial for
achieving this.

The presence of $\delta V$ leads to an additional contribution to the
isocurvature perturbations. On general grounds, one expects that
\be
\f{\delta\rho}{\rho}\Bigg\vert_{\delta V} 
\sim \m^2 \f{H_I}{2\pi f_I} \ .
\ee
CMB observations~\cite{Planck:2018jri} dictate that isocurvature
perturbations satisfy $\delta\rho_{\text{iso}}/\rho \lesssim 9\times 10^{-6}$
(see~\cite{Rigouzzo:2025hza}). Even if $H_I$ exceeds $f_I$ by many orders of
magnitude, this condition is automatically satisfied for all reasonable
parameters characterizing inflation and $f_a$, since $\mu\lesssim 10^{-10}$
due to the requirement of the smallness of the QCD $\theta$-angle, as shown
in \eq~(\ref{mubound}).

\paragraph*{Domain walls and their fate} 

At temperatures $T_{\rm osc}\sim 1$ GeV, corresponding
to~\cite{Sikivie:2006ni}
\be
t_{\rm osc}\sim 
2\times 10^{-7}\left(\f{f_a}{10^{12}~{\rm GeV}}\right)^\f{1}{3}~{\rm s} \ ,
\ee 
the QCD-induced axion mass is of the order of the Hubble rate,\footnote{Due
to the bounds~(\ref{mubound},\ref{DW}), the QCD-induced mass almost always
dominates over the explicit one. We discuss this in
appendix~\ref{app:onset}.} and the non-compact field creates a complicated
network of domain walls. This is very much different from the PQ scenario,
where walls are bounded by strings, and the number of their types (typically
of order one) is associated with the discrete unbroken subgroup of the U(1)
PQ symmetry. In our case, no strings are produced---there is no U(1) symmetry
to start with---while the number of (nearly) degenerate vacua within a
typical inflationary fluctuation largely exceeds one
\be
\label{eq:number}
N = \f{\delta\theta_I}{\delta
\theta_{\rm QCD}}\simeq \f{H_I}{f_I} \gg 1 \ .
\ee
This means that the adjacent ground states $n$ and $n+1$, with $n\lesssim N$,
are interpolated by  $\sim N$ different types of domain walls.

In the absence of $\delta V$ in~(\ref{eq:ALP_generic_Lagrangian}), these
configurations are stable, with approximate profile~\cite{Vilenkin:1982ks}
\be
\theta(z)=  4  {\rm arctan}\, e^{M_a z} \ ,
\ee
where $z$ is the coordinate perpendicular to their surface. Since their
energy density scales as $\rho_{\rm DW}\propto t^{-1}$, even one such defect
is enough to soon dominate in the Universe. This can well happen after the
time where BBN begins, $t_{\rm BBN}\sim 1~{\rm s}$, leading to catastrophic
consequences for cosmology~\cite{Zeldovich:1974uw}.

\begin{figure}[!t]
\centering 
\includegraphics[scale=.3]{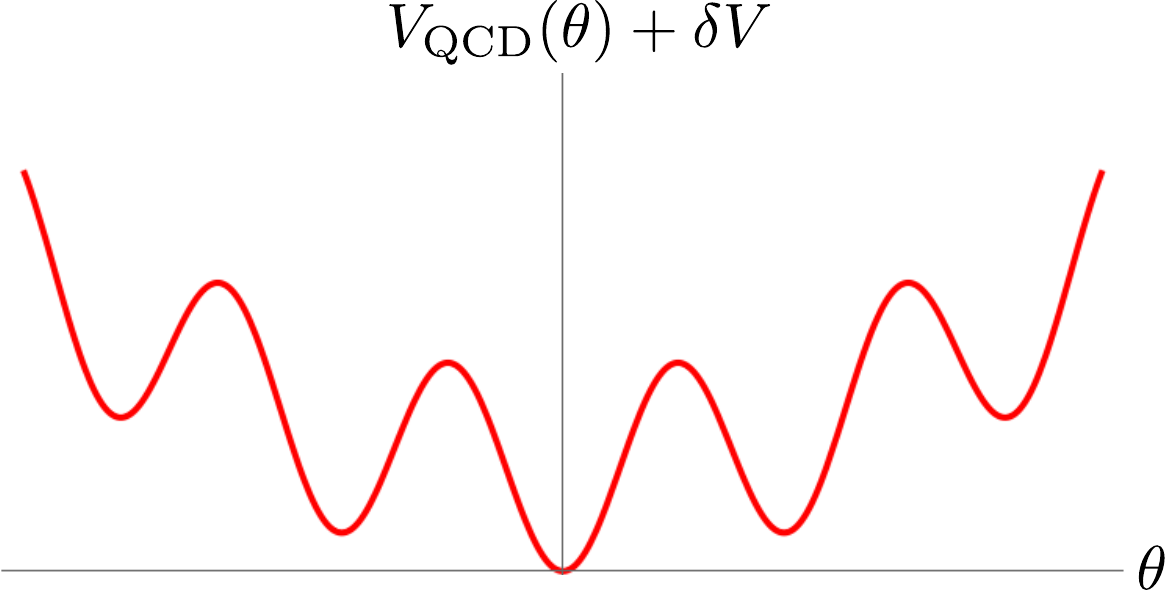}
\caption{The tilted potential, due to the presence of a small, non-QCD
related, mass for the axion.}
\label{fig:axion_potential_uplifted}
\end{figure}

The presence of $\delta V$, however, paints an entirely different
picture~\cite{Sikivie:1982qv}. As long as $m_a\ll M_a$,  the potential
retains an almost periodic structure, allowing domain walls to form; at the
same time, it can get sufficiently tilted to completely lift the degeneracy
of the ground states, as illustrated in
Fig.~\ref{fig:axion_potential_uplifted}.

Because of the difference in pressure on the two sides of each domain wall,
$\Delta V = 2\pi^2 (2n+1) m_a^2 f_a^2$, more energetically favorable vacua
will ``eat up'' the ones with larger energy densities. The instability
sets-in when $\Delta V$ exceeds the surface
tension~\cite{Sikivie:2006ni,GrillidiCortona:2015jxo} $\sigma_{\rm DW}\approx
9M_a f_a^2$, which happens around
\be
\label{eq:t_star}
t_\ast \sim 
\f{\sigma_{\rm DW}}{\Delta V} 
\sim \f{3}{(2n+1)\m^2} 
\left(\f{10^{-12}~{\rm GeV}}{M_a}\right)\times 10^{-13}~{\rm s} \ ,
\ee
where $\m$ was defined in~(\ref{eq:mass_ratio}). Note the $n$-dependence in
the above, in line with one's expectation that vacua with the largest $n$
collapse first, and those with $n\sim \mc O(1)$ last.

Requiring that the domain walls disappear before BBN, \ie $t_\ast\lesssim
1~{\rm s}$, we obtain
\be
\label{DW}
\m^2 \gtrsim  
\left(\f{10^{-12}~{\rm GeV}}{M_a}\right) \times 10^{-13} 
= 1.8\times10^{-11}\left(\frac{f_a}{10^{12}~{\rm GeV}}\right)\ . 
\ee
Because of eq.~(\ref{mubound}), this leads to a lower bound on the axion mass
$M_a>10^{-15}$ GeV and an upper bound on $f_a$, $f_a < 5.6\times 10^{12}$
GeV.

\paragraph*{Gravitational waves due to collapse of domain walls}

Following~\cite{Sikivie:1982qv}, assume that there is roughly one domain wall
per Hubble patch. During its shrinking, gravitational waves are radiated
away. Let us now present order-of-magnitude estimates for their typical
frequency and amplitude, as the analysis of the evolution of the domain wall
network would require numerical simulations (see
\eg~\cite{Saikawa:2017hiv,Dankovsky:2024zvs,Babichev:2025stm}) and goes
beyond the scope of the paper.

From~(\ref{eq:t_star}), we find that the characteristic frequency corresponds
to
\be
\label{eq:GW_frequency}
\omega_\ast \sim \f{1}{t_\ast} \sim  
\m^2~\left(\f{M_a}{10^{-12}~{\rm GeV}}\right)\times 10^{13}~{\rm s}^{-1} \ , 
\ee

Let us now make a rather crude estimate for the energy density $\Omega_{\rm
DW}$ of the produced gravitational waves. The amplitude of metric
perturbations sourced by the domain wall is
\be
h \sim \f{\sigma_{\rm DW}t}{M_{\rm Pl}^2} \ ,
\ee
with $M_{\rm Pl} = 2.44 \times 10^{18}~{\rm GeV}$, the Planck mass.
Gravitational radiation is due to the quadrupole moment of the collapsing
wall, therefore
\be
\Omega_{\rm GW} \propto M_{\rm Pl}^2\f{\dot h^2}{\rho} \ ,
\ee
where $\dot{}$ stands for derivative wrt time, and $\rho \propto M_{\rm
Pl}^2/t^2$, as follows from the Friedmann equation. From the above, and
accounting for numerical factors, we find
that~\cite{Dankovsky:2024zvs,Babichev:2025stm}
\be
\Omega_{\rm GW,\ast} \simeq 1.2\times 10^{-2}  
\left(\f{\sigma_{\rm DW} t_\ast}{M_{\rm Pl}^2}\right)^2 \ .
\ee
Using~(\ref{eq:t_star}), we can recast this as
\be
\label{eq:GW_amplitude}
\Omega_{\rm GW,\ast} \simeq 2\times 10^{-2} \m^{-4} 
\left(\f{f_a}{M_{\rm Pl}}\right)^4 \ .
\ee

Redshifting~(\ref{eq:GW_frequency},\ref{eq:GW_amplitude}) to the present day,
and using, e.g.~$\mu=10^{-6}$ and $f_a=3.2\times10^{11}~{\rm GeV}$ (c.f.~\eq
\eqref{eq:low_energy_fa}), yields the following peak frequency and amplitude
\be
\omega_0 \sim \mc O(1)~{\rm nHz} \ ,
~~~\Omega_{\rm GW,0} \sim \mc O(10^{-10}) \ .
\ee
This falls within the frequency band of pulsar timing
arrays~\cite{NANOGrav:2023gor,EPTA:2023fyk,Reardon:2023gzh,Xu:2023wog}, and
may be within reach of future observations~\cite{Babak:2024yhu}. Notice
though that the amplitude depends on the fourth power of the axionic decay
constant, meaning that a slight increase can have significant effects in
$\Omega_{\rm GW}$.

\paragraph*{Axion dark matter}

The ratio between $\Omega_\theta$ and the cold DM abundance $\Omega_{\rm DM}$
due to misalignment reads~\cite{ParticleDataGroup:2024cfk}
\be
\label{eq:Omega_DM}
\f{\Omega_\theta}{\Omega_{\rm DM}} \simeq 
\langle \delta\theta_I^2\rangle
\left(\f{f_a}{9\times 10^{11}~{\rm GeV}}\right)^{1.165} \ ,
\ee
where $\langle\delta\theta_I^2\rangle = \f{1}{2\pi}
\limitint^{+\pi}_{-\pi}\diff\theta\, \theta^2  = \f{\pi^2}{3} \ .$  We expect
that the annihilation of domain walls will also produce axions with an
abundance similar to eq.~(\ref{eq:Omega_DM}), but the precise contribution
from this process is unknown at present. The requirement
$\f{\Omega_\theta}{\Omega_{\rm DM}} <1$ leads to a constraint
\be
\label{eq:low_energy_fa}
f_a \lesssim \frac{3.2\times10^{11}~{\rm GeV} }{(1+\kappa)^\f{1}{1.165} }.
\ee
where the factor $\kappa \sim 1$ accounts for the domain wall source of axion
production. This is above the typical (model dependent) lower bound $f_a
\gtrsim \mc O(10^{9})~{\rm GeV}$ inferred from astrophysical
observations~\cite{Caputo:2024oqc} .

To account for all Dark Matter, $f_a$ should saturate the upper bound in
(\ref{eq:low_energy_fa}). If this is the case, the requirement (\ref{DW}) of
domain walls collapsing before BBN leads to a {\em lower} bound on $\mu$,
$\mu^2>4.3 \times 10^{-12}$. Since CP violation scales with $\mu$ as in
\eq~(\ref{CPViolation}), this scenario~\emph{predicts} that there {\em must
be} strong CP-violation not far from the present constraints on it:
\be
\theta_{\rm phys} \gtrsim 4.3\times 10^{-12} \ .
\ee
This is roughly 30 times smaller than the current bound~(\ref{eq:EDM_bound})
on the neutron electric dipole moment. The use of the astrophysical
constraint $f_a>10^{9}$ weakens the bound on $\mu$ by two orders of
magnitude, $\mu^2>2\times 10^{-14}$. In this case axions represent only 1\%
of DM, and the lower bound on $\theta_{\rm phys}$ is roughly four orders of
magnitude below the experimental constraint. In these estimates we neglected
$\kappa$.

It is intriguing that forthcoming measurements~\cite{Anastassopoulos:2015ura}
of the proton electric dipole moment are expected to probe strong CP
violation at an unprecedented level, improving the current
bound~(\ref{eq:EDM_bound}) by three orders of magnitude, and hence open a
direct experimental window on this proposal.

\paragraph*{Origin of non-compact axion}

The effective theory~(\ref{eq:ALP_generic_Lagrangian}) can have its roots in
top-down constructions linked to the nature of gravity. While in the simplest
constructions torsion cannot give rise to a viable QCD
axion~\cite{Karananas:2025ews}, an example that can fulfill the assumptions
made in the previous sections is the Weyl-invariant Einstein-Cartan
gravity~\cite{Karananas:2024xja,Karananas:2025ews}; we give details in
appendix~\ref{app:EinsteinCartan}.

First, the axion is deeply rooted in gravity, as it descends directly from
torsion. Thus, it is non-compact, which allows it to span the multiple
(would-be) minima of the QCD potential.

Second, and by construction, the axion has a field-dependent non-canonical
kinetic term. As for the coupling to QCD, controlled by $F$, it is generated
radiatively in a Weyl-invariant regularization
scheme~\cite{Shaposhnikov:2025znm,Karananas:2025ews}, so its exact form
cannot be unambiguously fixed from first principles.

Finally, the part of the axion potential which is not related to QCD
dynamics---here an explicit mass term---is in fact a necessary consistency
requirement~\cite{Karananas:2024xja,Karananas:2024hoh,Karananas:2024qrz,Karananas:2025ews}.
Importantly, it is controlled by one of the couplings of the gauged Lorentz
group $\widetilde g$ as
\be
m_a \propto \widetilde g M_{\rm Pl} \ .
\ee
The smallness of $m_a$ dictates that $\widetilde g\lll 1$, in full agreement
with what was anticipated in the aforementioned works. Therefore, what
initially appears as an ad-hoc phenomenological input preventing the domain
walls from overclosing the Universe, in this context is a manifestation of
the consistency of the gravitational dynamics.

\paragraph*{Conclusions and Outlook}

We have presented a cosmological axion framework in which the axion field is
fundamentally non-compact. In this setup, the axion can constitute a
substantial fraction---or the entirety---of the dark matter abundance.
Inflationary fluctuations populate many QCD vacua, thereby suppressing axion
isocurvature perturbations while unavoidably leading to the formation of
domain walls. A small non-QCD contribution to the axion potential lifts the
degeneracy among vacua, induces a residual amount of strong CP violation, and
triggers the collapse of the domain-wall network before Big Bang
Nucleosynthesis. The resulting annihilation of domain walls sources a
stochastic background of gravitational waves with characteristic frequencies
in the nanohertz range.

Taken together, these effects and the associated multiplicity of vacua give
rise to a correlated set of observational signatures: a nonzero strong CP
phase close to current experimental limits and a low-frequency
gravitational-wave background. Both signals lie below present sensitivities
but are potentially accessible to forthcoming electric dipole moment searches
and pulsar timing array experiments. Observation of either signal would
provide nontrivial information about the microscopic nature of the axion,
while their simultaneous detection would strongly point toward non-compact
axion dynamics. Such signatures would be absent if the axion is compact, or
if---according to
Refs.~\cite{Dvali:2013eja,Dvali:2014gua,Dvali:2017eba,Dvali:2018fqu,Dvali:2018jhn,Dvali:2018txx,Dvali:2018dce,Dvali:2022fdv}---meta-stable
de Sitter vacua are incompatible with quantum gravity. In this way, axion
phenomenology may offer a rare observational window into fundamental aspects
of quantum gravity.

A number of open issues remain and call for further study. In particular, our
estimates of the axion relic abundance and of the gravitational-wave spectrum
should be regarded as order-of-magnitude only. The production of axions
during domain-wall collapse, as well as the amplitude and spectral shape of
the emitted gravitational waves, depend sensitively on the detailed dynamics
of biased domain walls. Recent numerical studies~\cite{Babichev:2025stm}
indicate that the annihilation time of such networks---and hence the
resulting gravitational-wave signal---can differ significantly from analytic
expectations. A dedicated numerical analysis tailored to the present
framework is therefore required, and we leave this important task for future
work.

\begin{acknowledgments}

\paragraph*{Acknowledgments} 

The work of S.Z.~was supported by the European Research Council Gravites
Horizon Grant AO number: 850 173-6.

\textbf{Disclaimer:} Funded by the European Union. Views and opinions
 expressed are however those of the authors only and do not necessarily
 reflect those of the European Union or European Research Council. Neither
 the European Union nor the granting authority can be held responsible for
 them.
\end{acknowledgments}


\bibliography{Refs.bib}

\begin{thebibliography}{71}%
\makeatletter
\providecommand \@ifxundefined [1]{%
 \@ifx{#1\undefined}
}%
\providecommand \@ifnum [1]{%
 \ifnum #1\expandafter \@firstoftwo
 \else \expandafter \@secondoftwo
 \fi
}%
\providecommand \@ifx [1]{%
 \ifx #1\expandafter \@firstoftwo
 \else \expandafter \@secondoftwo
 \fi
}%
\providecommand \natexlab [1]{#1}%
\providecommand \enquote  [1]{``#1''}%
\providecommand \bibnamefont  [1]{#1}%
\providecommand \bibfnamefont [1]{#1}%
\providecommand \citenamefont [1]{#1}%
\providecommand \href@noop [0]{\@secondoftwo}%
\providecommand \href [0]{\begingroup \@sanitize@url \@href}%
\providecommand \@href[1]{\@@startlink{#1}\@@href}%
\providecommand \@@href[1]{\endgroup#1\@@endlink}%
\providecommand \@sanitize@url [0]{\catcode `\\12\catcode `\$12\catcode
  `\&12\catcode `\#12\catcode `\^12\catcode `\_12\catcode `\%12\relax}%
\providecommand \@@startlink[1]{}%
\providecommand \@@endlink[0]{}%
\providecommand \url  [0]{\begingroup\@sanitize@url \@url }%
\providecommand \@url [1]{\endgroup\@href {#1}{\urlprefix }}%
\providecommand \urlprefix  [0]{URL }%
\providecommand \Eprint [0]{\href }%
\providecommand \doibase [0]{https://doi.org/}%
\providecommand \selectlanguage [0]{\@gobble}%
\providecommand \bibinfo  [0]{\@secondoftwo}%
\providecommand \bibfield  [0]{\@secondoftwo}%
\providecommand \translation [1]{[#1]}%
\providecommand \BibitemOpen [0]{}%
\providecommand \bibitemStop [0]{}%
\providecommand \bibitemNoStop [0]{.\EOS\space}%
\providecommand \EOS [0]{\spacefactor3000\relax}%
\providecommand \BibitemShut  [1]{\csname bibitem#1\endcsname}%
\let\auto@bib@innerbib\@empty
\bibitem [{\citenamefont {Peccei}\ and\ \citenamefont
  {Quinn}(1977)}]{Peccei:1977hh}%
  \BibitemOpen
  \bibfield  {author} {\bibinfo {author} {\bibfnamefont {R.~D.}\ \bibnamefont
  {Peccei}}\ and\ \bibinfo {author} {\bibfnamefont {H.~R.}\ \bibnamefont
  {Quinn}},\ }\bibfield  {title} {\bibinfo {title} {{CP Conservation in the
  Presence of Instantons}},\ }\href
  {https://doi.org/10.1103/PhysRevLett.38.1440} {\bibfield  {journal} {\bibinfo
   {journal} {Phys. Rev. Lett.}\ }\textbf {\bibinfo {volume} {38}},\ \bibinfo
  {pages} {1440} (\bibinfo {year} {1977})}\BibitemShut {NoStop}%
\bibitem [{\citenamefont {Weinberg}(1978)}]{Weinberg:1977ma}%
  \BibitemOpen
  \bibfield  {author} {\bibinfo {author} {\bibfnamefont {S.}~\bibnamefont
  {Weinberg}},\ }\bibfield  {title} {\bibinfo {title} {{A New Light Boson?}},\
  }\href {https://doi.org/10.1103/PhysRevLett.40.223} {\bibfield  {journal}
  {\bibinfo  {journal} {Phys. Rev. Lett.}\ }\textbf {\bibinfo {volume} {40}},\
  \bibinfo {pages} {223} (\bibinfo {year} {1978})}\BibitemShut {NoStop}%
\bibitem [{\citenamefont {Wilczek}(1978)}]{Wilczek:1977pj}%
  \BibitemOpen
  \bibfield  {author} {\bibinfo {author} {\bibfnamefont {F.}~\bibnamefont
  {Wilczek}},\ }\bibfield  {title} {\bibinfo {title} {{Problem of Strong $P$
  and $T$ Invariance in the Presence of Instantons}},\ }\href
  {https://doi.org/10.1103/PhysRevLett.40.279} {\bibfield  {journal} {\bibinfo
  {journal} {Phys. Rev. Lett.}\ }\textbf {\bibinfo {volume} {40}},\ \bibinfo
  {pages} {279} (\bibinfo {year} {1978})}\BibitemShut {NoStop}%
\bibitem [{\citenamefont {Witten}(1984)}]{Witten:1984dg}%
  \BibitemOpen
  \bibfield  {author} {\bibinfo {author} {\bibfnamefont {E.}~\bibnamefont
  {Witten}},\ }\bibfield  {title} {\bibinfo {title} {{Some Properties of O(32)
  Superstrings}},\ }\href {https://doi.org/10.1016/0370-2693(84)90422-2}
  {\bibfield  {journal} {\bibinfo  {journal} {Phys. Lett. B}\ }\textbf
  {\bibinfo {volume} {149}},\ \bibinfo {pages} {351} (\bibinfo {year}
  {1984})}\BibitemShut {NoStop}%
\bibitem [{\citenamefont {Reece}(2025)}]{Reece:2024wrn}%
  \BibitemOpen
  \bibfield  {author} {\bibinfo {author} {\bibfnamefont {M.}~\bibnamefont
  {Reece}},\ }\bibfield  {title} {\bibinfo {title} {{Extra-dimensional axion
  expectations}},\ }\href {https://doi.org/10.1007/JHEP07(2025)130} {\bibfield
  {journal} {\bibinfo  {journal} {JHEP}\ }\textbf {\bibinfo {volume} {07}},\
  \bibinfo {pages} {130}},\ \Eprint {https://arxiv.org/abs/2406.08543}
  {arXiv:2406.08543 [hep-ph]} \BibitemShut {NoStop}%
\bibitem [{\citenamefont {Dvali}(2005)}]{Dvali:2005an}%
  \BibitemOpen
  \bibfield  {author} {\bibinfo {author} {\bibfnamefont {G.}~\bibnamefont
  {Dvali}},\ }\bibfield  {title} {\bibinfo {title} {{Three-form gauging of
  axion symmetries and gravity}},\ }\href@noop {} {\  (\bibinfo {year}
  {2005})},\ \Eprint {https://arxiv.org/abs/hep-th/0507215}
  {arXiv:hep-th/0507215} \BibitemShut {NoStop}%
\bibitem [{\citenamefont {Dvali}\ \emph {et~al.}(2014)\citenamefont {Dvali},
  \citenamefont {Folkerts},\ and\ \citenamefont {Franca}}]{Dvali:2013cpa}%
  \BibitemOpen
  \bibfield  {author} {\bibinfo {author} {\bibfnamefont {G.}~\bibnamefont
  {Dvali}}, \bibinfo {author} {\bibfnamefont {S.}~\bibnamefont {Folkerts}},\
  and\ \bibinfo {author} {\bibfnamefont {A.}~\bibnamefont {Franca}},\
  }\bibfield  {title} {\bibinfo {title} {{How neutrino protects the axion}},\
  }\href {https://doi.org/10.1103/PhysRevD.89.105025} {\bibfield  {journal}
  {\bibinfo  {journal} {Phys. Rev. D}\ }\textbf {\bibinfo {volume} {89}},\
  \bibinfo {pages} {105025} (\bibinfo {year} {2014})},\ \Eprint
  {https://arxiv.org/abs/1312.7273} {arXiv:1312.7273 [hep-th]} \BibitemShut
  {NoStop}%
\bibitem [{\citenamefont {Dvali}(2017)}]{Dvali:2017mpy}%
  \BibitemOpen
  \bibfield  {author} {\bibinfo {author} {\bibfnamefont {G.}~\bibnamefont
  {Dvali}},\ }\bibfield  {title} {\bibinfo {title} {{Topological Origin of
  Chiral Symmetry Breaking in QCD and in Gravity}},\ }\href@noop {} {\
  (\bibinfo {year} {2017})},\ \Eprint {https://arxiv.org/abs/1705.06317}
  {arXiv:1705.06317 [hep-th]} \BibitemShut {NoStop}%
\bibitem [{\citenamefont {Dvali}(2022)}]{Dvali:2022fdv}%
  \BibitemOpen
  \bibfield  {author} {\bibinfo {author} {\bibfnamefont {G.}~\bibnamefont
  {Dvali}},\ }\bibfield  {title} {\bibinfo {title} {{Strong-$CP$ with and
  without gravity}},\ }\href@noop {} {\  (\bibinfo {year} {2022})},\ \Eprint
  {https://arxiv.org/abs/2209.14219} {arXiv:2209.14219 [hep-ph]} \BibitemShut
  {NoStop}%
\bibitem [{\citenamefont {Mercuri}(2009)}]{Mercuri:2009zi}%
  \BibitemOpen
  \bibfield  {author} {\bibinfo {author} {\bibfnamefont {S.}~\bibnamefont
  {Mercuri}},\ }\bibfield  {title} {\bibinfo {title} {{Peccei-Quinn mechanism
  in gravity and the nature of the Barbero-Immirzi parameter}},\ }\href
  {https://doi.org/10.1103/PhysRevLett.103.081302} {\bibfield  {journal}
  {\bibinfo  {journal} {Phys. Rev. Lett.}\ }\textbf {\bibinfo {volume} {103}},\
  \bibinfo {pages} {081302} (\bibinfo {year} {2009})},\ \Eprint
  {https://arxiv.org/abs/0902.2764} {arXiv:0902.2764 [gr-qc]} \BibitemShut
  {NoStop}%
\bibitem [{\citenamefont {Karananas}\ \emph {et~al.}(2024)\citenamefont
  {Karananas}, \citenamefont {Shaposhnikov},\ and\ \citenamefont
  {Zell}}]{Karananas:2024xja}%
  \BibitemOpen
  \bibfield  {author} {\bibinfo {author} {\bibfnamefont {G.~K.}\ \bibnamefont
  {Karananas}}, \bibinfo {author} {\bibfnamefont {M.}~\bibnamefont
  {Shaposhnikov}},\ and\ \bibinfo {author} {\bibfnamefont {S.}~\bibnamefont
  {Zell}},\ }\bibfield  {title} {\bibinfo {title} {{Weyl-invariant
  Einstein-Cartan gravity: unifying the strong CP and hierarchy puzzles}},\
  }\href {https://doi.org/10.1007/JHEP11(2024)146} {\bibfield  {journal}
  {\bibinfo  {journal} {JHEP}\ }\textbf {\bibinfo {volume} {11}},\ \bibinfo
  {pages} {146}},\ \Eprint {https://arxiv.org/abs/2406.11956} {arXiv:2406.11956
  [hep-th]} \BibitemShut {NoStop}%
\bibitem [{\citenamefont {Karananas}\ \emph
  {et~al.}(2025{\natexlab{a}})\citenamefont {Karananas}, \citenamefont
  {Shaposhnikov},\ and\ \citenamefont {Zell}}]{Karananas:2025ews}%
  \BibitemOpen
  \bibfield  {author} {\bibinfo {author} {\bibfnamefont {G.~K.}\ \bibnamefont
  {Karananas}}, \bibinfo {author} {\bibfnamefont {M.}~\bibnamefont
  {Shaposhnikov}},\ and\ \bibinfo {author} {\bibfnamefont {S.}~\bibnamefont
  {Zell}},\ }\bibfield  {title} {\bibinfo {title} {{Gravitational Origin of the
  QCD Axion}},\ }\href {https://doi.org/10.1103/d7mp-sjvc} {\bibfield
  {journal} {\bibinfo  {journal} {Phys. Rev. Lett.}\ }\textbf {\bibinfo
  {volume} {135}},\ \bibinfo {pages} {241001} (\bibinfo {year}
  {2025}{\natexlab{a}})},\ \Eprint {https://arxiv.org/abs/2506.11836}
  {arXiv:2506.11836 [hep-th]} \BibitemShut {NoStop}%
\bibitem [{\citenamefont {Marsh}(2016)}]{Marsh:2015xka}%
  \BibitemOpen
  \bibfield  {author} {\bibinfo {author} {\bibfnamefont {D.~J.~E.}\
  \bibnamefont {Marsh}},\ }\bibfield  {title} {\bibinfo {title} {{Axion
  Cosmology}},\ }\href {https://doi.org/10.1016/j.physrep.2016.06.005}
  {\bibfield  {journal} {\bibinfo  {journal} {Phys. Rept.}\ }\textbf {\bibinfo
  {volume} {643}},\ \bibinfo {pages} {1} (\bibinfo {year} {2016})},\ \Eprint
  {https://arxiv.org/abs/1510.07633} {arXiv:1510.07633 [astro-ph.CO]}
  \BibitemShut {NoStop}%
\bibitem [{\citenamefont {Di~Luzio}\ \emph {et~al.}(2020)\citenamefont
  {Di~Luzio}, \citenamefont {Giannotti}, \citenamefont {Nardi},\ and\
  \citenamefont {Visinelli}}]{DiLuzio:2020wdo}%
  \BibitemOpen
  \bibfield  {author} {\bibinfo {author} {\bibfnamefont {L.}~\bibnamefont
  {Di~Luzio}}, \bibinfo {author} {\bibfnamefont {M.}~\bibnamefont {Giannotti}},
  \bibinfo {author} {\bibfnamefont {E.}~\bibnamefont {Nardi}},\ and\ \bibinfo
  {author} {\bibfnamefont {L.}~\bibnamefont {Visinelli}},\ }\bibfield  {title}
  {\bibinfo {title} {{The landscape of QCD axion models}},\ }\href
  {https://doi.org/10.1016/j.physrep.2020.06.002} {\bibfield  {journal}
  {\bibinfo  {journal} {Phys. Rept.}\ }\textbf {\bibinfo {volume} {870}},\
  \bibinfo {pages} {1} (\bibinfo {year} {2020})},\ \Eprint
  {https://arxiv.org/abs/2003.01100} {arXiv:2003.01100 [hep-ph]} \BibitemShut
  {NoStop}%
\bibitem [{\citenamefont {Choi}\ \emph {et~al.}(2021)\citenamefont {Choi},
  \citenamefont {Im},\ and\ \citenamefont {Sub~Shin}}]{Choi:2020rgn}%
  \BibitemOpen
  \bibfield  {author} {\bibinfo {author} {\bibfnamefont {K.}~\bibnamefont
  {Choi}}, \bibinfo {author} {\bibfnamefont {S.~H.}\ \bibnamefont {Im}},\ and\
  \bibinfo {author} {\bibfnamefont {C.}~\bibnamefont {Sub~Shin}},\ }\bibfield
  {title} {\bibinfo {title} {{Recent Progress in the Physics of Axions and
  Axion-Like Particles}},\ }\href
  {https://doi.org/10.1146/annurev-nucl-120720-031147} {\bibfield  {journal}
  {\bibinfo  {journal} {Ann. Rev. Nucl. Part. Sci.}\ }\textbf {\bibinfo
  {volume} {71}},\ \bibinfo {pages} {225} (\bibinfo {year} {2021})},\ \Eprint
  {https://arxiv.org/abs/2012.05029} {arXiv:2012.05029 [hep-ph]} \BibitemShut
  {NoStop}%
\bibitem [{\citenamefont {Turner}\ and\ \citenamefont
  {Wilczek}(1991)}]{Turner:1990uz}%
  \BibitemOpen
  \bibfield  {author} {\bibinfo {author} {\bibfnamefont {M.~S.}\ \bibnamefont
  {Turner}}\ and\ \bibinfo {author} {\bibfnamefont {F.}~\bibnamefont
  {Wilczek}},\ }\bibfield  {title} {\bibinfo {title} {{Inflationary Axion
  Cosmology}},\ }\href {https://doi.org/10.1103/PhysRevLett.66.5} {\bibfield
  {journal} {\bibinfo  {journal} {Phys. Rev. Lett.}\ }\textbf {\bibinfo
  {volume} {66}},\ \bibinfo {pages} {5} (\bibinfo {year} {1991})}\BibitemShut
  {NoStop}%
\bibitem [{\citenamefont {Akrami}\ \emph {et~al.}(2020)\citenamefont {Akrami}
  \emph {et~al.}}]{Planck:2018jri}%
  \BibitemOpen
  \bibfield  {author} {\bibinfo {author} {\bibfnamefont {Y.}~\bibnamefont
  {Akrami}} \emph {et~al.} (\bibinfo {collaboration} {Planck}),\ }\bibfield
  {title} {\bibinfo {title} {{Planck 2018 results. X. Constraints on
  inflation}},\ }\href {https://doi.org/10.1051/0004-6361/201833887} {\bibfield
   {journal} {\bibinfo  {journal} {Astron. Astrophys.}\ }\textbf {\bibinfo
  {volume} {641}},\ \bibinfo {pages} {A10} (\bibinfo {year} {2020})},\ \Eprint
  {https://arxiv.org/abs/1807.06211} {arXiv:1807.06211 [astro-ph.CO]}
  \BibitemShut {NoStop}%
\bibitem [{\citenamefont {Kofman}(1986)}]{Kofman:1985zx}%
  \BibitemOpen
  \bibfield  {author} {\bibinfo {author} {\bibfnamefont {L.~A.}\ \bibnamefont
  {Kofman}},\ }\bibfield  {title} {\bibinfo {title} {{What Initial
  Perturbations May Be Generated in Inflationary Cosmological Models}},\ }\href
  {https://doi.org/10.1016/0370-2693(86)90403-X} {\bibfield  {journal}
  {\bibinfo  {journal} {Phys. Lett. B}\ }\textbf {\bibinfo {volume} {173}},\
  \bibinfo {pages} {400} (\bibinfo {year} {1986})}\BibitemShut {NoStop}%
\bibitem [{\citenamefont {Kofman}\ and\ \citenamefont
  {Linde}(1987)}]{Kofman:1986wm}%
  \BibitemOpen
  \bibfield  {author} {\bibinfo {author} {\bibfnamefont {L.~A.}\ \bibnamefont
  {Kofman}}\ and\ \bibinfo {author} {\bibfnamefont {A.~D.}\ \bibnamefont
  {Linde}},\ }\bibfield  {title} {\bibinfo {title} {{Generation of Density
  Perturbations in the Inflationary Cosmology}},\ }\href
  {https://doi.org/10.1016/0550-3213(87)90698-5} {\bibfield  {journal}
  {\bibinfo  {journal} {Nucl. Phys. B}\ }\textbf {\bibinfo {volume} {282}},\
  \bibinfo {pages} {555} (\bibinfo {year} {1987})}\BibitemShut {NoStop}%
\bibitem [{\citenamefont {Linde}\ and\ \citenamefont
  {Lyth}(1990)}]{Linde:1990yj}%
  \BibitemOpen
  \bibfield  {author} {\bibinfo {author} {\bibfnamefont {A.~D.}\ \bibnamefont
  {Linde}}\ and\ \bibinfo {author} {\bibfnamefont {D.~H.}\ \bibnamefont
  {Lyth}},\ }\bibfield  {title} {\bibinfo {title} {{Axionic domain wall
  production during inflation}},\ }\href
  {https://doi.org/10.1016/0370-2693(90)90613-B} {\bibfield  {journal}
  {\bibinfo  {journal} {Phys. Lett. B}\ }\textbf {\bibinfo {volume} {246}},\
  \bibinfo {pages} {353} (\bibinfo {year} {1990})}\BibitemShut {NoStop}%
\bibitem [{\citenamefont {Sikivie}(1982)}]{Sikivie:1982qv}%
  \BibitemOpen
  \bibfield  {author} {\bibinfo {author} {\bibfnamefont {P.}~\bibnamefont
  {Sikivie}},\ }\bibfield  {title} {\bibinfo {title} {{Of Axions, Domain Walls
  and the Early Universe}},\ }\href
  {https://doi.org/10.1103/PhysRevLett.48.1156} {\bibfield  {journal} {\bibinfo
   {journal} {Phys. Rev. Lett.}\ }\textbf {\bibinfo {volume} {48}},\ \bibinfo
  {pages} {1156} (\bibinfo {year} {1982})}\BibitemShut {NoStop}%
\bibitem [{\citenamefont {Abel}\ \emph {et~al.}(2020)\citenamefont {Abel} \emph
  {et~al.}}]{Abel:2020pzs}%
  \BibitemOpen
  \bibfield  {author} {\bibinfo {author} {\bibfnamefont {C.}~\bibnamefont
  {Abel}} \emph {et~al.},\ }\bibfield  {title} {\bibinfo {title} {{Measurement
  of the Permanent Electric Dipole Moment of the Neutron}},\ }\href
  {https://doi.org/10.1103/PhysRevLett.124.081803} {\bibfield  {journal}
  {\bibinfo  {journal} {Phys. Rev. Lett.}\ }\textbf {\bibinfo {volume} {124}},\
  \bibinfo {pages} {081803} (\bibinfo {year} {2020})},\ \Eprint
  {https://arxiv.org/abs/2001.11966} {arXiv:2001.11966 [hep-ex]} \BibitemShut
  {NoStop}%
\bibitem [{\citenamefont {Anastassopoulos}\ \emph {et~al.}(2016)\citenamefont
  {Anastassopoulos} \emph {et~al.}}]{Anastassopoulos:2015ura}%
  \BibitemOpen
  \bibfield  {author} {\bibinfo {author} {\bibfnamefont {V.}~\bibnamefont
  {Anastassopoulos}} \emph {et~al.},\ }\bibfield  {title} {\bibinfo {title} {{A
  Storage Ring Experiment to Detect a Proton Electric Dipole Moment}},\ }\href
  {https://doi.org/10.1063/1.4967465} {\bibfield  {journal} {\bibinfo
  {journal} {Rev. Sci. Instrum.}\ }\textbf {\bibinfo {volume} {87}},\ \bibinfo
  {pages} {115116} (\bibinfo {year} {2016})},\ \Eprint
  {https://arxiv.org/abs/1502.04317} {arXiv:1502.04317 [physics.acc-ph]}
  \BibitemShut {NoStop}%
\bibitem [{\citenamefont {Kawasaki}\ \emph {et~al.}(2013)\citenamefont
  {Kawasaki}, \citenamefont {Yanagida},\ and\ \citenamefont
  {Yoshino}}]{Kawasaki:2013iha}%
  \BibitemOpen
  \bibfield  {author} {\bibinfo {author} {\bibfnamefont {M.}~\bibnamefont
  {Kawasaki}}, \bibinfo {author} {\bibfnamefont {T.~T.}\ \bibnamefont
  {Yanagida}},\ and\ \bibinfo {author} {\bibfnamefont {K.}~\bibnamefont
  {Yoshino}},\ }\bibfield  {title} {\bibinfo {title} {{Domain wall and
  isocurvature perturbation problems in axion models}},\ }\href
  {https://doi.org/10.1088/1475-7516/2013/11/030} {\bibfield  {journal}
  {\bibinfo  {journal} {JCAP}\ }\textbf {\bibinfo {volume} {11}},\ \bibinfo
  {pages} {030}},\ \Eprint {https://arxiv.org/abs/1305.5338} {arXiv:1305.5338
  [hep-ph]} \BibitemShut {NoStop}%
\bibitem [{\citenamefont {Harigaya}\ \emph {et~al.}(2015)\citenamefont
  {Harigaya}, \citenamefont {Ibe}, \citenamefont {Kawasaki},\ and\
  \citenamefont {Yanagida}}]{Harigaya:2015hha}%
  \BibitemOpen
  \bibfield  {author} {\bibinfo {author} {\bibfnamefont {K.}~\bibnamefont
  {Harigaya}}, \bibinfo {author} {\bibfnamefont {M.}~\bibnamefont {Ibe}},
  \bibinfo {author} {\bibfnamefont {M.}~\bibnamefont {Kawasaki}},\ and\
  \bibinfo {author} {\bibfnamefont {T.~T.}\ \bibnamefont {Yanagida}},\
  }\bibfield  {title} {\bibinfo {title} {{Dynamics of Peccei-Quinn Breaking
  Field after Inflation and Axion Isocurvature Perturbations}},\ }\href
  {https://doi.org/10.1088/1475-7516/2015/11/003} {\bibfield  {journal}
  {\bibinfo  {journal} {JCAP}\ }\textbf {\bibinfo {volume} {11}},\ \bibinfo
  {pages} {003}},\ \Eprint {https://arxiv.org/abs/1507.00119} {arXiv:1507.00119
  [hep-ph]} \BibitemShut {NoStop}%
\bibitem [{\citenamefont {Kearney}\ \emph {et~al.}(2016)\citenamefont
  {Kearney}, \citenamefont {Orlofsky},\ and\ \citenamefont
  {Pierce}}]{Kearney:2016vqw}%
  \BibitemOpen
  \bibfield  {author} {\bibinfo {author} {\bibfnamefont {J.}~\bibnamefont
  {Kearney}}, \bibinfo {author} {\bibfnamefont {N.}~\bibnamefont {Orlofsky}},\
  and\ \bibinfo {author} {\bibfnamefont {A.}~\bibnamefont {Pierce}},\
  }\bibfield  {title} {\bibinfo {title} {{High-Scale Axions without
  Isocurvature from Inflationary Dynamics}},\ }\href
  {https://doi.org/10.1103/PhysRevD.93.095026} {\bibfield  {journal} {\bibinfo
  {journal} {Phys. Rev. D}\ }\textbf {\bibinfo {volume} {93}},\ \bibinfo
  {pages} {095026} (\bibinfo {year} {2016})},\ \Eprint
  {https://arxiv.org/abs/1601.03049} {arXiv:1601.03049 [hep-ph]} \BibitemShut
  {NoStop}%
\bibitem [{\citenamefont {Kobayashi}\ and\ \citenamefont
  {Takahashi}(2016)}]{Kobayashi:2016qld}%
  \BibitemOpen
  \bibfield  {author} {\bibinfo {author} {\bibfnamefont {T.}~\bibnamefont
  {Kobayashi}}\ and\ \bibinfo {author} {\bibfnamefont {F.}~\bibnamefont
  {Takahashi}},\ }\bibfield  {title} {\bibinfo {title} {{Cosmological
  Perturbations of Axion with a Dynamical Decay Constant}},\ }\href
  {https://doi.org/10.1088/1475-7516/2016/08/056} {\bibfield  {journal}
  {\bibinfo  {journal} {JCAP}\ }\textbf {\bibinfo {volume} {08}},\ \bibinfo
  {pages} {056}},\ \Eprint {https://arxiv.org/abs/1607.04294} {arXiv:1607.04294
  [hep-ph]} \BibitemShut {NoStop}%
\bibitem [{\citenamefont {Co}\ \emph {et~al.}(2018)\citenamefont {Co},
  \citenamefont {Hall},\ and\ \citenamefont {Harigaya}}]{Co:2017mop}%
  \BibitemOpen
  \bibfield  {author} {\bibinfo {author} {\bibfnamefont {R.~T.}\ \bibnamefont
  {Co}}, \bibinfo {author} {\bibfnamefont {L.~J.}\ \bibnamefont {Hall}},\ and\
  \bibinfo {author} {\bibfnamefont {K.}~\bibnamefont {Harigaya}},\ }\bibfield
  {title} {\bibinfo {title} {{QCD Axion Dark Matter with a Small Decay
  Constant}},\ }\href {https://doi.org/10.1103/PhysRevLett.120.211602}
  {\bibfield  {journal} {\bibinfo  {journal} {Phys. Rev. Lett.}\ }\textbf
  {\bibinfo {volume} {120}},\ \bibinfo {pages} {211602} (\bibinfo {year}
  {2018})},\ \Eprint {https://arxiv.org/abs/1711.10486} {arXiv:1711.10486
  [hep-ph]} \BibitemShut {NoStop}%
\bibitem [{\citenamefont {Co}\ \emph {et~al.}(2020)\citenamefont {Co},
  \citenamefont {Hall}, \citenamefont {Harigaya}, \citenamefont {Olive},\ and\
  \citenamefont {Verner}}]{Co:2020dya}%
  \BibitemOpen
  \bibfield  {author} {\bibinfo {author} {\bibfnamefont {R.~T.}\ \bibnamefont
  {Co}}, \bibinfo {author} {\bibfnamefont {L.~J.}\ \bibnamefont {Hall}},
  \bibinfo {author} {\bibfnamefont {K.}~\bibnamefont {Harigaya}}, \bibinfo
  {author} {\bibfnamefont {K.~A.}\ \bibnamefont {Olive}},\ and\ \bibinfo
  {author} {\bibfnamefont {S.}~\bibnamefont {Verner}},\ }\bibfield  {title}
  {\bibinfo {title} {{Axion Kinetic Misalignment and Parametric Resonance from
  Inflation}},\ }\href {https://doi.org/10.1088/1475-7516/2020/08/036}
  {\bibfield  {journal} {\bibinfo  {journal} {JCAP}\ }\textbf {\bibinfo
  {volume} {08}},\ \bibinfo {pages} {036}},\ \Eprint
  {https://arxiv.org/abs/2004.00629} {arXiv:2004.00629 [hep-ph]} \BibitemShut
  {NoStop}%
\bibitem [{\citenamefont {Ballesteros}\ \emph {et~al.}(2021)\citenamefont
  {Ballesteros}, \citenamefont {Ringwald}, \citenamefont {Tamarit},\ and\
  \citenamefont {Welling}}]{Ballesteros:2021bee}%
  \BibitemOpen
  \bibfield  {author} {\bibinfo {author} {\bibfnamefont {G.}~\bibnamefont
  {Ballesteros}}, \bibinfo {author} {\bibfnamefont {A.}~\bibnamefont
  {Ringwald}}, \bibinfo {author} {\bibfnamefont {C.}~\bibnamefont {Tamarit}},\
  and\ \bibinfo {author} {\bibfnamefont {Y.}~\bibnamefont {Welling}},\
  }\bibfield  {title} {\bibinfo {title} {{Revisiting isocurvature bounds in
  models unifying the axion with the inflaton}},\ }\href
  {https://doi.org/10.1088/1475-7516/2021/09/036} {\bibfield  {journal}
  {\bibinfo  {journal} {JCAP}\ }\textbf {\bibinfo {volume} {09}},\ \bibinfo
  {pages} {036}},\ \Eprint {https://arxiv.org/abs/2104.13847} {arXiv:2104.13847
  [hep-ph]} \BibitemShut {NoStop}%
\bibitem [{\citenamefont {Graham}\ and\ \citenamefont
  {Racco}(2025)}]{Graham:2025iwx}%
  \BibitemOpen
  \bibfield  {author} {\bibinfo {author} {\bibfnamefont {P.~W.}\ \bibnamefont
  {Graham}}\ and\ \bibinfo {author} {\bibfnamefont {D.}~\bibnamefont {Racco}},\
  }\bibfield  {title} {\bibinfo {title} {{Revisiting isocurvature bounds on the
  minimal QCD axion}},\ }\href {https://doi.org/10.1007/JHEP12(2025)028}
  {\bibfield  {journal} {\bibinfo  {journal} {JHEP}\ }\textbf {\bibinfo
  {volume} {12}},\ \bibinfo {pages} {028}},\ \Eprint
  {https://arxiv.org/abs/2506.03348} {arXiv:2506.03348 [hep-ph]} \BibitemShut
  {NoStop}%
\bibitem [{\citenamefont {Rigouzzo}\ and\ \citenamefont
  {Zell}(2025{\natexlab{a}})}]{Rigouzzo:2025hza}%
  \BibitemOpen
  \bibfield  {author} {\bibinfo {author} {\bibfnamefont {C.}~\bibnamefont
  {Rigouzzo}}\ and\ \bibinfo {author} {\bibfnamefont {S.}~\bibnamefont
  {Zell}},\ }\bibfield  {title} {\bibinfo {title} {{No Dark Matter Axion During
  Minimal Higgs Inflation}},\ }\href@noop {} {\  (\bibinfo {year}
  {2025}{\natexlab{a}})},\ \Eprint {https://arxiv.org/abs/2504.02952}
  {arXiv:2504.02952 [hep-ph]} \BibitemShut {NoStop}%
\bibitem [{\citenamefont {Rigouzzo}\ and\ \citenamefont
  {Zell}(2025{\natexlab{b}})}]{Rigouzzo:2025ycb}%
  \BibitemOpen
  \bibfield  {author} {\bibinfo {author} {\bibfnamefont {C.}~\bibnamefont
  {Rigouzzo}}\ and\ \bibinfo {author} {\bibfnamefont {S.}~\bibnamefont
  {Zell}},\ }\bibfield  {title} {\bibinfo {title} {{On Non-Minimal Couplings to
  Gravity and Axion Isocurvature Bounds}},\ }\href@noop {} {\  (\bibinfo {year}
  {2025}{\natexlab{b}})},\ \Eprint {https://arxiv.org/abs/2512.16754}
  {arXiv:2512.16754 [hep-ph]} \BibitemShut {NoStop}%
\bibitem [{\citenamefont {Garcia-Bellido}\ \emph {et~al.}(2012)\citenamefont
  {Garcia-Bellido}, \citenamefont {Rubio},\ and\ \citenamefont
  {Shaposhnikov}}]{Garcia-Bellido:2012npk}%
  \BibitemOpen
  \bibfield  {author} {\bibinfo {author} {\bibfnamefont {J.}~\bibnamefont
  {Garcia-Bellido}}, \bibinfo {author} {\bibfnamefont {J.}~\bibnamefont
  {Rubio}},\ and\ \bibinfo {author} {\bibfnamefont {M.}~\bibnamefont
  {Shaposhnikov}},\ }\bibfield  {title} {\bibinfo {title} {{Higgs-Dilaton
  cosmology: Are there extra relativistic species?}},\ }\href
  {https://doi.org/10.1016/j.physletb.2012.10.075} {\bibfield  {journal}
  {\bibinfo  {journal} {Phys. Lett. B}\ }\textbf {\bibinfo {volume} {718}},\
  \bibinfo {pages} {507} (\bibinfo {year} {2012})},\ \Eprint
  {https://arxiv.org/abs/1209.2119} {arXiv:1209.2119 [hep-ph]} \BibitemShut
  {NoStop}%
\bibitem [{\citenamefont {Navas}\ \emph {et~al.}(2024)\citenamefont {Navas}
  \emph {et~al.}}]{ParticleDataGroup:2024cfk}%
  \BibitemOpen
  \bibfield  {author} {\bibinfo {author} {\bibfnamefont {S.}~\bibnamefont
  {Navas}} \emph {et~al.} (\bibinfo {collaboration} {Particle Data Group}),\
  }\bibfield  {title} {\bibinfo {title} {{Review of particle physics}},\ }\href
  {https://doi.org/10.1103/PhysRevD.110.030001} {\bibfield  {journal} {\bibinfo
   {journal} {Phys. Rev. D}\ }\textbf {\bibinfo {volume} {110}},\ \bibinfo
  {pages} {030001} (\bibinfo {year} {2024})}\BibitemShut {NoStop}%
\bibitem [{\citenamefont {Bezrukov}\ and\ \citenamefont
  {Shaposhnikov}(2008)}]{Bezrukov:2007ep}%
  \BibitemOpen
  \bibfield  {author} {\bibinfo {author} {\bibfnamefont {F.~L.}\ \bibnamefont
  {Bezrukov}}\ and\ \bibinfo {author} {\bibfnamefont {M.}~\bibnamefont
  {Shaposhnikov}},\ }\bibfield  {title} {\bibinfo {title} {{The Standard Model
  Higgs boson as the inflaton}},\ }\href
  {https://doi.org/10.1016/j.physletb.2007.11.072} {\bibfield  {journal}
  {\bibinfo  {journal} {Phys. Lett. B}\ }\textbf {\bibinfo {volume} {659}},\
  \bibinfo {pages} {703} (\bibinfo {year} {2008})},\ \Eprint
  {https://arxiv.org/abs/0710.3755} {arXiv:0710.3755 [hep-th]} \BibitemShut
  {NoStop}%
\bibitem [{\citenamefont {Zell}(2025)}]{Zell:2024cyz}%
  \BibitemOpen
  \bibfield  {author} {\bibinfo {author} {\bibfnamefont {S.}~\bibnamefont
  {Zell}},\ }\bibfield  {title} {\bibinfo {title} {{No warm inflation from a
  vanilla axion}},\ }\href {https://doi.org/10.1103/kp15-3s1c} {\bibfield
  {journal} {\bibinfo  {journal} {Phys. Rev. D}\ }\textbf {\bibinfo {volume}
  {112}},\ \bibinfo {pages} {L081307} (\bibinfo {year} {2025})},\ \Eprint
  {https://arxiv.org/abs/2408.07746} {arXiv:2408.07746 [hep-ph]} \BibitemShut
  {NoStop}%
\bibitem [{\citenamefont {Berghaus}\ \emph {et~al.}(2025)\citenamefont
  {Berghaus}, \citenamefont {Drewes},\ and\ \citenamefont
  {Zell}}]{Berghaus:2025dqi}%
  \BibitemOpen
  \bibfield  {author} {\bibinfo {author} {\bibfnamefont {K.~V.}\ \bibnamefont
  {Berghaus}}, \bibinfo {author} {\bibfnamefont {M.}~\bibnamefont {Drewes}},\
  and\ \bibinfo {author} {\bibfnamefont {S.}~\bibnamefont {Zell}},\ }\bibfield
  {title} {\bibinfo {title} {{Warm Inflation with the Standard Model}},\ }\href
  {https://doi.org/10.1103/9nn9-bsm9} {\bibfield  {journal} {\bibinfo
  {journal} {Phys. Rev. Lett.}\ }\textbf {\bibinfo {volume} {135}},\ \bibinfo
  {pages} {171002} (\bibinfo {year} {2025})},\ \Eprint
  {https://arxiv.org/abs/2503.18829} {arXiv:2503.18829 [hep-ph]} \BibitemShut
  {NoStop}%
\bibitem [{\citenamefont {Sikivie}(2008)}]{Sikivie:2006ni}%
  \BibitemOpen
  \bibfield  {author} {\bibinfo {author} {\bibfnamefont {P.}~\bibnamefont
  {Sikivie}},\ }\bibfield  {title} {\bibinfo {title} {{Axion Cosmology}},\
  }\href {https://doi.org/10.1007/978-3-540-73518-2_2} {\bibfield  {journal}
  {\bibinfo  {journal} {Lect. Notes Phys.}\ }\textbf {\bibinfo {volume}
  {741}},\ \bibinfo {pages} {19} (\bibinfo {year} {2008})},\ \Eprint
  {https://arxiv.org/abs/astro-ph/0610440} {arXiv:astro-ph/0610440}
  \BibitemShut {NoStop}%
\bibitem [{\citenamefont {Vilenkin}\ and\ \citenamefont
  {Everett}(1982)}]{Vilenkin:1982ks}%
  \BibitemOpen
  \bibfield  {author} {\bibinfo {author} {\bibfnamefont {A.}~\bibnamefont
  {Vilenkin}}\ and\ \bibinfo {author} {\bibfnamefont {A.~E.}\ \bibnamefont
  {Everett}},\ }\bibfield  {title} {\bibinfo {title} {{Cosmic Strings and
  Domain Walls in Models with Goldstone and PseudoGoldstone Bosons}},\ }\href
  {https://doi.org/10.1103/PhysRevLett.48.1867} {\bibfield  {journal} {\bibinfo
   {journal} {Phys. Rev. Lett.}\ }\textbf {\bibinfo {volume} {48}},\ \bibinfo
  {pages} {1867} (\bibinfo {year} {1982})}\BibitemShut {NoStop}%
\bibitem [{\citenamefont {Zeldovich}\ \emph {et~al.}(1974)\citenamefont
  {Zeldovich}, \citenamefont {Kobzarev},\ and\ \citenamefont
  {Okun}}]{Zeldovich:1974uw}%
  \BibitemOpen
  \bibfield  {author} {\bibinfo {author} {\bibfnamefont {Y.~B.}\ \bibnamefont
  {Zeldovich}}, \bibinfo {author} {\bibfnamefont {I.~Y.}\ \bibnamefont
  {Kobzarev}},\ and\ \bibinfo {author} {\bibfnamefont {L.~B.}\ \bibnamefont
  {Okun}},\ }\bibfield  {title} {\bibinfo {title} {{Cosmological Consequences
  of the Spontaneous Breakdown of Discrete Symmetry}},\ }\href@noop {}
  {\bibfield  {journal} {\bibinfo  {journal} {Zh. Eksp. Teor. Fiz.}\ }\textbf
  {\bibinfo {volume} {67}},\ \bibinfo {pages} {3} (\bibinfo {year}
  {1974})}\BibitemShut {NoStop}%
\bibitem [{\citenamefont {Grilli~di Cortona}\ \emph {et~al.}(2016)\citenamefont
  {Grilli~di Cortona}, \citenamefont {Hardy}, \citenamefont {Pardo~Vega},\ and\
  \citenamefont {Villadoro}}]{GrillidiCortona:2015jxo}%
  \BibitemOpen
  \bibfield  {author} {\bibinfo {author} {\bibfnamefont {G.}~\bibnamefont
  {Grilli~di Cortona}}, \bibinfo {author} {\bibfnamefont {E.}~\bibnamefont
  {Hardy}}, \bibinfo {author} {\bibfnamefont {J.}~\bibnamefont {Pardo~Vega}},\
  and\ \bibinfo {author} {\bibfnamefont {G.}~\bibnamefont {Villadoro}},\
  }\bibfield  {title} {\bibinfo {title} {{The QCD axion, precisely}},\ }\href
  {https://doi.org/10.1007/JHEP01(2016)034} {\bibfield  {journal} {\bibinfo
  {journal} {JHEP}\ }\textbf {\bibinfo {volume} {01}},\ \bibinfo {pages}
  {034}},\ \Eprint {https://arxiv.org/abs/1511.02867} {arXiv:1511.02867
  [hep-ph]} \BibitemShut {NoStop}%
\bibitem [{\citenamefont {Saikawa}(2017)}]{Saikawa:2017hiv}%
  \BibitemOpen
  \bibfield  {author} {\bibinfo {author} {\bibfnamefont {K.}~\bibnamefont
  {Saikawa}},\ }\bibfield  {title} {\bibinfo {title} {{A review of
  gravitational waves from cosmic domain walls}},\ }\href
  {https://doi.org/10.3390/universe3020040} {\bibfield  {journal} {\bibinfo
  {journal} {Universe}\ }\textbf {\bibinfo {volume} {3}},\ \bibinfo {pages}
  {40} (\bibinfo {year} {2017})},\ \Eprint {https://arxiv.org/abs/1703.02576}
  {arXiv:1703.02576 [hep-ph]} \BibitemShut {NoStop}%
\bibitem [{\citenamefont {Dankovsky}\ \emph {et~al.}(2024)\citenamefont
  {Dankovsky}, \citenamefont {Babichev}, \citenamefont {Gorbunov},
  \citenamefont {Ramazanov},\ and\ \citenamefont {Vikman}}]{Dankovsky:2024zvs}%
  \BibitemOpen
  \bibfield  {author} {\bibinfo {author} {\bibfnamefont {I.}~\bibnamefont
  {Dankovsky}}, \bibinfo {author} {\bibfnamefont {E.}~\bibnamefont {Babichev}},
  \bibinfo {author} {\bibfnamefont {D.}~\bibnamefont {Gorbunov}}, \bibinfo
  {author} {\bibfnamefont {S.}~\bibnamefont {Ramazanov}},\ and\ \bibinfo
  {author} {\bibfnamefont {A.}~\bibnamefont {Vikman}},\ }\bibfield  {title}
  {\bibinfo {title} {{Revisiting evolution of domain walls and their
  gravitational radiation with CosmoLattice}},\ }\href
  {https://doi.org/10.1088/1475-7516/2024/09/047} {\bibfield  {journal}
  {\bibinfo  {journal} {JCAP}\ }\textbf {\bibinfo {volume} {09}},\ \bibinfo
  {pages} {047}},\ \Eprint {https://arxiv.org/abs/2406.17053} {arXiv:2406.17053
  [astro-ph.CO]} \BibitemShut {NoStop}%
\bibitem [{\citenamefont {Babichev}\ \emph {et~al.}(2025)\citenamefont
  {Babichev}, \citenamefont {Dankovsky}, \citenamefont {Gorbunov},
  \citenamefont {Ramazanov},\ and\ \citenamefont {Vikman}}]{Babichev:2025stm}%
  \BibitemOpen
  \bibfield  {author} {\bibinfo {author} {\bibfnamefont {E.}~\bibnamefont
  {Babichev}}, \bibinfo {author} {\bibfnamefont {I.}~\bibnamefont {Dankovsky}},
  \bibinfo {author} {\bibfnamefont {D.}~\bibnamefont {Gorbunov}}, \bibinfo
  {author} {\bibfnamefont {S.}~\bibnamefont {Ramazanov}},\ and\ \bibinfo
  {author} {\bibfnamefont {A.}~\bibnamefont {Vikman}},\ }\bibfield  {title}
  {\bibinfo {title} {{Biased domain walls: faster annihilation, weaker
  gravitational waves}},\ }\href
  {https://doi.org/10.1088/1475-7516/2025/10/103} {\bibfield  {journal}
  {\bibinfo  {journal} {JCAP}\ }\textbf {\bibinfo {volume} {10}},\ \bibinfo
  {pages} {103}},\ \Eprint {https://arxiv.org/abs/2504.07902} {arXiv:2504.07902
  [hep-ph]} \BibitemShut {NoStop}%
\bibitem [{\citenamefont {Agazie}\ \emph {et~al.}(2023)\citenamefont {Agazie}
  \emph {et~al.}}]{NANOGrav:2023gor}%
  \BibitemOpen
  \bibfield  {author} {\bibinfo {author} {\bibfnamefont {G.}~\bibnamefont
  {Agazie}} \emph {et~al.} (\bibinfo {collaboration} {NANOGrav}),\ }\bibfield
  {title} {\bibinfo {title} {{The NANOGrav 15 yr Data Set: Evidence for a
  Gravitational-wave Background}},\ }\href
  {https://doi.org/10.3847/2041-8213/acdac6} {\bibfield  {journal} {\bibinfo
  {journal} {Astrophys. J. Lett.}\ }\textbf {\bibinfo {volume} {951}},\
  \bibinfo {pages} {L8} (\bibinfo {year} {2023})},\ \Eprint
  {https://arxiv.org/abs/2306.16213} {arXiv:2306.16213 [astro-ph.HE]}
  \BibitemShut {NoStop}%
\bibitem [{\citenamefont {Antoniadis}\ \emph {et~al.}(2023)\citenamefont
  {Antoniadis} \emph {et~al.}}]{EPTA:2023fyk}%
  \BibitemOpen
  \bibfield  {author} {\bibinfo {author} {\bibfnamefont {J.}~\bibnamefont
  {Antoniadis}} \emph {et~al.} (\bibinfo {collaboration} {EPTA, InPTA:}),\
  }\bibfield  {title} {\bibinfo {title} {{The second data release from the
  European Pulsar Timing Array - III. Search for gravitational wave signals}},\
  }\href {https://doi.org/10.1051/0004-6361/202346844} {\bibfield  {journal}
  {\bibinfo  {journal} {Astron. Astrophys.}\ }\textbf {\bibinfo {volume}
  {678}},\ \bibinfo {pages} {A50} (\bibinfo {year} {2023})},\ \Eprint
  {https://arxiv.org/abs/2306.16214} {arXiv:2306.16214 [astro-ph.HE]}
  \BibitemShut {NoStop}%
\bibitem [{\citenamefont {Reardon}\ \emph {et~al.}(2023)\citenamefont {Reardon}
  \emph {et~al.}}]{Reardon:2023gzh}%
  \BibitemOpen
  \bibfield  {author} {\bibinfo {author} {\bibfnamefont {D.~J.}\ \bibnamefont
  {Reardon}} \emph {et~al.},\ }\bibfield  {title} {\bibinfo {title} {{Search
  for an Isotropic Gravitational-wave Background with the Parkes Pulsar Timing
  Array}},\ }\href {https://doi.org/10.3847/2041-8213/acdd02} {\bibfield
  {journal} {\bibinfo  {journal} {Astrophys. J. Lett.}\ }\textbf {\bibinfo
  {volume} {951}},\ \bibinfo {pages} {L6} (\bibinfo {year} {2023})},\ \Eprint
  {https://arxiv.org/abs/2306.16215} {arXiv:2306.16215 [astro-ph.HE]}
  \BibitemShut {NoStop}%
\bibitem [{\citenamefont {Xu}\ \emph {et~al.}(2023)\citenamefont {Xu} \emph
  {et~al.}}]{Xu:2023wog}%
  \BibitemOpen
  \bibfield  {author} {\bibinfo {author} {\bibfnamefont {H.}~\bibnamefont {Xu}}
  \emph {et~al.},\ }\bibfield  {title} {\bibinfo {title} {{Searching for the
  Nano-Hertz Stochastic Gravitational Wave Background with the Chinese Pulsar
  Timing Array Data Release I}},\ }\href
  {https://doi.org/10.1088/1674-4527/acdfa5} {\bibfield  {journal} {\bibinfo
  {journal} {Res. Astron. Astrophys.}\ }\textbf {\bibinfo {volume} {23}},\
  \bibinfo {pages} {075024} (\bibinfo {year} {2023})},\ \Eprint
  {https://arxiv.org/abs/2306.16216} {arXiv:2306.16216 [astro-ph.HE]}
  \BibitemShut {NoStop}%
\bibitem [{\citenamefont {Babak}\ \emph {et~al.}(2024)\citenamefont {Babak},
  \citenamefont {Falxa}, \citenamefont {Franciolini},\ and\ \citenamefont
  {Pieroni}}]{Babak:2024yhu}%
  \BibitemOpen
  \bibfield  {author} {\bibinfo {author} {\bibfnamefont {S.}~\bibnamefont
  {Babak}}, \bibinfo {author} {\bibfnamefont {M.}~\bibnamefont {Falxa}},
  \bibinfo {author} {\bibfnamefont {G.}~\bibnamefont {Franciolini}},\ and\
  \bibinfo {author} {\bibfnamefont {M.}~\bibnamefont {Pieroni}},\ }\bibfield
  {title} {\bibinfo {title} {{Forecasting the sensitivity of pulsar timing
  arrays to gravitational wave backgrounds}},\ }\href
  {https://doi.org/10.1103/PhysRevD.110.063022} {\bibfield  {journal} {\bibinfo
   {journal} {Phys. Rev. D}\ }\textbf {\bibinfo {volume} {110}},\ \bibinfo
  {pages} {063022} (\bibinfo {year} {2024})},\ \Eprint
  {https://arxiv.org/abs/2404.02864} {arXiv:2404.02864 [astro-ph.CO]}
  \BibitemShut {NoStop}%
\bibitem [{\citenamefont {Caputo}\ and\ \citenamefont
  {Raffelt}(2024)}]{Caputo:2024oqc}%
  \BibitemOpen
  \bibfield  {author} {\bibinfo {author} {\bibfnamefont {A.}~\bibnamefont
  {Caputo}}\ and\ \bibinfo {author} {\bibfnamefont {G.}~\bibnamefont
  {Raffelt}},\ }\bibfield  {title} {\bibinfo {title} {{Astrophysical Axion
  Bounds: The 2024 Edition}},\ }\href {https://doi.org/10.22323/1.454.0041}
  {\bibfield  {journal} {\bibinfo  {journal} {PoS}\ }\textbf {\bibinfo {volume}
  {COSMICWISPers}},\ \bibinfo {pages} {041} (\bibinfo {year} {2024})},\ \Eprint
  {https://arxiv.org/abs/2401.13728} {arXiv:2401.13728 [hep-ph]} \BibitemShut
  {NoStop}%
\bibitem [{\citenamefont {Shaposhnikov}(2025)}]{Shaposhnikov:2025znm}%
  \BibitemOpen
  \bibfield  {author} {\bibinfo {author} {\bibfnamefont {M.}~\bibnamefont
  {Shaposhnikov}},\ }\bibinfo {title} {{Progress in Einstein-Cartan gravity}}\
  (\bibinfo {year} {2025})\ \Eprint {https://arxiv.org/abs/2506.11847}
  {arXiv:2506.11847 [hep-th]} \BibitemShut {NoStop}%
\bibitem [{\citenamefont {Karananas}(2025)}]{Karananas:2024hoh}%
  \BibitemOpen
  \bibfield  {author} {\bibinfo {author} {\bibfnamefont {G.~K.}\ \bibnamefont
  {Karananas}},\ }\bibfield  {title} {\bibinfo {title} {{Particle content of
  (scalar{\,}curvature)2 gravities revisited}},\ }\href
  {https://doi.org/10.1103/PhysRevD.111.044068} {\bibfield  {journal} {\bibinfo
   {journal} {Phys. Rev. D}\ }\textbf {\bibinfo {volume} {111}},\ \bibinfo
  {pages} {044068} (\bibinfo {year} {2025})},\ \Eprint
  {https://arxiv.org/abs/2407.09598} {arXiv:2407.09598 [hep-th]} \BibitemShut
  {NoStop}%
\bibitem [{\citenamefont {Karananas}(2024)}]{Karananas:2024qrz}%
  \BibitemOpen
  \bibfield  {author} {\bibinfo {author} {\bibfnamefont {G.~K.}\ \bibnamefont
  {Karananas}},\ }\bibfield  {title} {\bibinfo {title} {{The particle content
  of (scalar curvature)$^2$ metric-affine gravity}},\ }\href@noop {} {\
  (\bibinfo {year} {2024})},\ \Eprint {https://arxiv.org/abs/2408.16818}
  {arXiv:2408.16818 [hep-th]} \BibitemShut {NoStop}%
\bibitem [{\citenamefont {Dvali}\ and\ \citenamefont
  {Gomez}(2014)}]{Dvali:2013eja}%
  \BibitemOpen
  \bibfield  {author} {\bibinfo {author} {\bibfnamefont {G.}~\bibnamefont
  {Dvali}}\ and\ \bibinfo {author} {\bibfnamefont {C.}~\bibnamefont {Gomez}},\
  }\bibfield  {title} {\bibinfo {title} {{Quantum Compositeness of Gravity:
  Black Holes, AdS and Inflation}},\ }\href
  {https://doi.org/10.1088/1475-7516/2014/01/023} {\bibfield  {journal}
  {\bibinfo  {journal} {JCAP}\ }\textbf {\bibinfo {volume} {01}},\ \bibinfo
  {pages} {023}},\ \Eprint {https://arxiv.org/abs/1312.4795} {arXiv:1312.4795
  [hep-th]} \BibitemShut {NoStop}%
\bibitem [{\citenamefont {Dvali}\ and\ \citenamefont
  {Gomez}(2016)}]{Dvali:2014gua}%
  \BibitemOpen
  \bibfield  {author} {\bibinfo {author} {\bibfnamefont {G.}~\bibnamefont
  {Dvali}}\ and\ \bibinfo {author} {\bibfnamefont {C.}~\bibnamefont {Gomez}},\
  }\bibfield  {title} {\bibinfo {title} {{Quantum Exclusion of Positive
  Cosmological Constant?}},\ }\href {https://doi.org/10.1002/andp.201500216}
  {\bibfield  {journal} {\bibinfo  {journal} {Annalen Phys.}\ }\textbf
  {\bibinfo {volume} {528}},\ \bibinfo {pages} {68} (\bibinfo {year} {2016})},\
  \Eprint {https://arxiv.org/abs/1412.8077} {arXiv:1412.8077 [hep-th]}
  \BibitemShut {NoStop}%
\bibitem [{\citenamefont {Dvali}\ \emph {et~al.}(2017)\citenamefont {Dvali},
  \citenamefont {Gomez},\ and\ \citenamefont {Zell}}]{Dvali:2017eba}%
  \BibitemOpen
  \bibfield  {author} {\bibinfo {author} {\bibfnamefont {G.}~\bibnamefont
  {Dvali}}, \bibinfo {author} {\bibfnamefont {C.}~\bibnamefont {Gomez}},\ and\
  \bibinfo {author} {\bibfnamefont {S.}~\bibnamefont {Zell}},\ }\bibfield
  {title} {\bibinfo {title} {{Quantum Break-Time of de Sitter}},\ }\href
  {https://doi.org/10.1088/1475-7516/2017/06/028} {\bibfield  {journal}
  {\bibinfo  {journal} {JCAP}\ }\textbf {\bibinfo {volume} {06}},\ \bibinfo
  {pages} {028}},\ \Eprint {https://arxiv.org/abs/1701.08776} {arXiv:1701.08776
  [hep-th]} \BibitemShut {NoStop}%
\bibitem [{\citenamefont {Dvali}\ and\ \citenamefont
  {Gomez}(2019)}]{Dvali:2018fqu}%
  \BibitemOpen
  \bibfield  {author} {\bibinfo {author} {\bibfnamefont {G.}~\bibnamefont
  {Dvali}}\ and\ \bibinfo {author} {\bibfnamefont {C.}~\bibnamefont {Gomez}},\
  }\bibfield  {title} {\bibinfo {title} {{On Exclusion of Positive Cosmological
  Constant}},\ }\href {https://doi.org/10.1002/prop.201800092} {\bibfield
  {journal} {\bibinfo  {journal} {Fortsch. Phys.}\ }\textbf {\bibinfo {volume}
  {67}},\ \bibinfo {pages} {1800092} (\bibinfo {year} {2019})},\ \Eprint
  {https://arxiv.org/abs/1806.10877} {arXiv:1806.10877 [hep-th]} \BibitemShut
  {NoStop}%
\bibitem [{\citenamefont {Dvali}\ \emph {et~al.}(2019)\citenamefont {Dvali},
  \citenamefont {Gomez},\ and\ \citenamefont {Zell}}]{Dvali:2018jhn}%
  \BibitemOpen
  \bibfield  {author} {\bibinfo {author} {\bibfnamefont {G.}~\bibnamefont
  {Dvali}}, \bibinfo {author} {\bibfnamefont {C.}~\bibnamefont {Gomez}},\ and\
  \bibinfo {author} {\bibfnamefont {S.}~\bibnamefont {Zell}},\ }\bibfield
  {title} {\bibinfo {title} {{Quantum Breaking Bound on de Sitter and
  Swampland}},\ }\href {https://doi.org/10.1002/prop.201800094} {\bibfield
  {journal} {\bibinfo  {journal} {Fortsch. Phys.}\ }\textbf {\bibinfo {volume}
  {67}},\ \bibinfo {pages} {1800094} (\bibinfo {year} {2019})},\ \Eprint
  {https://arxiv.org/abs/1810.11002} {arXiv:1810.11002 [hep-th]} \BibitemShut
  {NoStop}%
\bibitem [{\citenamefont {Dvali}\ \emph
  {et~al.}(2018{\natexlab{a}})\citenamefont {Dvali}, \citenamefont {Gomez},\
  and\ \citenamefont {Zell}}]{Dvali:2018txx}%
  \BibitemOpen
  \bibfield  {author} {\bibinfo {author} {\bibfnamefont {G.}~\bibnamefont
  {Dvali}}, \bibinfo {author} {\bibfnamefont {C.}~\bibnamefont {Gomez}},\ and\
  \bibinfo {author} {\bibfnamefont {S.}~\bibnamefont {Zell}},\ }\bibfield
  {title} {\bibinfo {title} {{Discrete Symmetries Excluded by Quantum
  Breaking}},\ }\href@noop {} {\  (\bibinfo {year} {2018}{\natexlab{a}})},\
  \Eprint {https://arxiv.org/abs/1811.03077} {arXiv:1811.03077 [hep-th]}
  \BibitemShut {NoStop}%
\bibitem [{\citenamefont {Dvali}\ \emph
  {et~al.}(2018{\natexlab{b}})\citenamefont {Dvali}, \citenamefont {Gomez},\
  and\ \citenamefont {Zell}}]{Dvali:2018dce}%
  \BibitemOpen
  \bibfield  {author} {\bibinfo {author} {\bibfnamefont {G.}~\bibnamefont
  {Dvali}}, \bibinfo {author} {\bibfnamefont {C.}~\bibnamefont {Gomez}},\ and\
  \bibinfo {author} {\bibfnamefont {S.}~\bibnamefont {Zell}},\ }\bibfield
  {title} {\bibinfo {title} {{A Proof of the Axion?}},\ }\href@noop {} {\
  (\bibinfo {year} {2018}{\natexlab{b}})},\ \Eprint
  {https://arxiv.org/abs/1811.03079} {arXiv:1811.03079 [hep-th]} \BibitemShut
  {NoStop}%
\bibitem [{\citenamefont {Karananas}\ and\ \citenamefont
  {Shaposhnikov}(2025)}]{Karananas:2025qsm}%
  \BibitemOpen
  \bibfield  {author} {\bibinfo {author} {\bibfnamefont {G.~K.}\ \bibnamefont
  {Karananas}}\ and\ \bibinfo {author} {\bibfnamefont {M.}~\bibnamefont
  {Shaposhnikov}},\ }\bibfield  {title} {\bibinfo {title} {{Higgs inflation in
  Weyl-invariant Einstein-Cartan gravity}},\ }\href@noop {} {\  (\bibinfo
  {year} {2025})},\ \Eprint {https://arxiv.org/abs/2511.04732}
  {arXiv:2511.04732 [hep-ph]} \BibitemShut {NoStop}%
\bibitem [{\citenamefont {Graf}\ and\ \citenamefont
  {Steffen}(2011)}]{Graf:2010tv}%
  \BibitemOpen
  \bibfield  {author} {\bibinfo {author} {\bibfnamefont {P.}~\bibnamefont
  {Graf}}\ and\ \bibinfo {author} {\bibfnamefont {F.~D.}\ \bibnamefont
  {Steffen}},\ }\bibfield  {title} {\bibinfo {title} {{Thermal axion production
  in the primordial quark-gluon plasma}},\ }\href
  {https://doi.org/10.1103/PhysRevD.83.075011} {\bibfield  {journal} {\bibinfo
  {journal} {Phys. Rev. D}\ }\textbf {\bibinfo {volume} {83}},\ \bibinfo
  {pages} {075011} (\bibinfo {year} {2011})},\ \Eprint
  {https://arxiv.org/abs/1008.4528} {arXiv:1008.4528 [hep-ph]} \BibitemShut
  {NoStop}%
\bibitem [{\citenamefont {Salvio}\ \emph {et~al.}(2014)\citenamefont {Salvio},
  \citenamefont {Strumia},\ and\ \citenamefont {Xue}}]{Salvio:2013iaa}%
  \BibitemOpen
  \bibfield  {author} {\bibinfo {author} {\bibfnamefont {A.}~\bibnamefont
  {Salvio}}, \bibinfo {author} {\bibfnamefont {A.}~\bibnamefont {Strumia}},\
  and\ \bibinfo {author} {\bibfnamefont {W.}~\bibnamefont {Xue}},\ }\bibfield
  {title} {\bibinfo {title} {{Thermal axion production}},\ }\href
  {https://doi.org/10.1088/1475-7516/2014/01/011} {\bibfield  {journal}
  {\bibinfo  {journal} {JCAP}\ }\textbf {\bibinfo {volume} {01}},\ \bibinfo
  {pages} {011}},\ \Eprint {https://arxiv.org/abs/1310.6982} {arXiv:1310.6982
  [hep-ph]} \BibitemShut {NoStop}%
\bibitem [{\citenamefont {D'Eramo}\ \emph {et~al.}(2021)\citenamefont
  {D'Eramo}, \citenamefont {Hajkarim},\ and\ \citenamefont
  {Yun}}]{DEramo:2021lgb}%
  \BibitemOpen
  \bibfield  {author} {\bibinfo {author} {\bibfnamefont {F.}~\bibnamefont
  {D'Eramo}}, \bibinfo {author} {\bibfnamefont {F.}~\bibnamefont {Hajkarim}},\
  and\ \bibinfo {author} {\bibfnamefont {S.}~\bibnamefont {Yun}},\ }\bibfield
  {title} {\bibinfo {title} {{Thermal QCD Axions across Thresholds}},\ }\href
  {https://doi.org/10.1007/JHEP10(2021)224} {\bibfield  {journal} {\bibinfo
  {journal} {JHEP}\ }\textbf {\bibinfo {volume} {10}},\ \bibinfo {pages}
  {224}},\ \Eprint {https://arxiv.org/abs/2108.05371} {arXiv:2108.05371
  [hep-ph]} \BibitemShut {NoStop}%
\bibitem [{\citenamefont {O'Hare}(2024)}]{OHare:2024nmr}%
  \BibitemOpen
  \bibfield  {author} {\bibinfo {author} {\bibfnamefont {C.~A.~J.}\
  \bibnamefont {O'Hare}},\ }\bibfield  {title} {\bibinfo {title} {{Cosmology of
  axion dark matter}},\ }\href {https://doi.org/10.22323/1.454.0040} {\bibfield
   {journal} {\bibinfo  {journal} {PoS}\ }\textbf {\bibinfo {volume}
  {COSMICWISPers}},\ \bibinfo {pages} {040} (\bibinfo {year} {2024})},\ \Eprint
  {https://arxiv.org/abs/2403.17697} {arXiv:2403.17697 [hep-ph]} \BibitemShut
  {NoStop}%
\bibitem [{\citenamefont {Karananas}\ \emph
  {et~al.}(2025{\natexlab{b}})\citenamefont {Karananas}, \citenamefont
  {Shaposhnikov},\ and\ \citenamefont {Zell}}]{Karananas:2025fas}%
  \BibitemOpen
  \bibfield  {author} {\bibinfo {author} {\bibfnamefont {G.~K.}\ \bibnamefont
  {Karananas}}, \bibinfo {author} {\bibfnamefont {M.}~\bibnamefont
  {Shaposhnikov}},\ and\ \bibinfo {author} {\bibfnamefont {S.}~\bibnamefont
  {Zell}},\ }\bibfield  {title} {\bibinfo {title} {{Weyl-invariant
  Einstein-Cartan gravity with a heavy ALP: Higgs Inflation and
  $\alpha$-attractors}},\ }\href@noop {} {\  (\bibinfo {year}
  {2025}{\natexlab{b}})},\ \Eprint {https://arxiv.org/abs/2507.15927}
  {arXiv:2507.15927 [hep-ph]} \BibitemShut {NoStop}%
\bibitem [{\citenamefont {Englert}\ \emph {et~al.}(1976)\citenamefont
  {Englert}, \citenamefont {Truffin},\ and\ \citenamefont
  {Gastmans}}]{Englert:1976ep}%
  \BibitemOpen
  \bibfield  {author} {\bibinfo {author} {\bibfnamefont {F.}~\bibnamefont
  {Englert}}, \bibinfo {author} {\bibfnamefont {C.}~\bibnamefont {Truffin}},\
  and\ \bibinfo {author} {\bibfnamefont {R.}~\bibnamefont {Gastmans}},\
  }\bibfield  {title} {\bibinfo {title} {{Conformal Invariance in Quantum
  Gravity}},\ }\href {https://doi.org/10.1016/0550-3213(76)90406-5} {\bibfield
  {journal} {\bibinfo  {journal} {Nucl. Phys. B}\ }\textbf {\bibinfo {volume}
  {117}},\ \bibinfo {pages} {407} (\bibinfo {year} {1976})}\BibitemShut
  {NoStop}%
\bibitem [{\citenamefont {Shaposhnikov}\ and\ \citenamefont
  {Zenhausern}(2009)}]{Shaposhnikov:2008xi}%
  \BibitemOpen
  \bibfield  {author} {\bibinfo {author} {\bibfnamefont {M.}~\bibnamefont
  {Shaposhnikov}}\ and\ \bibinfo {author} {\bibfnamefont {D.}~\bibnamefont
  {Zenhausern}},\ }\bibfield  {title} {\bibinfo {title} {{Quantum scale
  invariance, cosmological constant and hierarchy problem}},\ }\href
  {https://doi.org/10.1016/j.physletb.2008.11.041} {\bibfield  {journal}
  {\bibinfo  {journal} {Phys. Lett. B}\ }\textbf {\bibinfo {volume} {671}},\
  \bibinfo {pages} {162} (\bibinfo {year} {2009})},\ \Eprint
  {https://arxiv.org/abs/0809.3406} {arXiv:0809.3406 [hep-th]} \BibitemShut
  {NoStop}%
\bibitem [{\citenamefont {Shaposhnikov}\ and\ \citenamefont
  {Tkachov}(2009)}]{Shaposhnikov:2009nk}%
  \BibitemOpen
  \bibfield  {author} {\bibinfo {author} {\bibfnamefont {M.~E.}\ \bibnamefont
  {Shaposhnikov}}\ and\ \bibinfo {author} {\bibfnamefont {F.~V.}\ \bibnamefont
  {Tkachov}},\ }\bibfield  {title} {\bibinfo {title} {{Quantum scale-invariant
  models as effective field theories}},\ }\href@noop {} {\  (\bibinfo {year}
  {2009})},\ \Eprint {https://arxiv.org/abs/0905.4857} {arXiv:0905.4857
  [hep-th]} \BibitemShut {NoStop}%
\bibitem [{\citenamefont {Bezrukov}\ \emph {et~al.}(2011)\citenamefont
  {Bezrukov}, \citenamefont {Magnin}, \citenamefont {Shaposhnikov},\ and\
  \citenamefont {Sibiryakov}}]{Bezrukov:2010jz}%
  \BibitemOpen
  \bibfield  {author} {\bibinfo {author} {\bibfnamefont {F.}~\bibnamefont
  {Bezrukov}}, \bibinfo {author} {\bibfnamefont {A.}~\bibnamefont {Magnin}},
  \bibinfo {author} {\bibfnamefont {M.}~\bibnamefont {Shaposhnikov}},\ and\
  \bibinfo {author} {\bibfnamefont {S.}~\bibnamefont {Sibiryakov}},\ }\bibfield
   {title} {\bibinfo {title} {{Higgs inflation: consistency and
  generalisations}},\ }\href {https://doi.org/10.1007/JHEP01(2011)016}
  {\bibfield  {journal} {\bibinfo  {journal} {JHEP}\ }\textbf {\bibinfo
  {volume} {01}},\ \bibinfo {pages} {016}},\ \Eprint
  {https://arxiv.org/abs/1008.5157} {arXiv:1008.5157 [hep-ph]} \BibitemShut
  {NoStop}%
\end{thebibliography}%

\clearpage
\onecolumngrid

\appendix

\section{Self-consistency of the effective theory}
\label{app:effectiveTheory}

Our starting point is
\be
\label{eq:appendix_EFT_Lagrangian}
\mc L = -\f{f_I^2}{2}(\p_\m\theta)^2 
+ \f{g_s^2\theta}{32\pi^2 F_I}G^b_{\m\n}\widetilde G^{b\,\m\n} - \delta V \ ,
\ee
\ie~\eq~(\ref{eq:ALP_generic_Lagrangian}) of the main text taken at
inflation. In what follows we neglect $\delta V$ as it is irrelevant for the
following discussions.

It is convenient at this point to normalize canonically the field. Remember
though, $f_I$ is not constant; for simplicity we shall take it here to depend
on the inflaton, say $h$, only. Introducing
\be
\label{eq:appendix_canonical_axion}
a = f_I \theta \ ,
\ee
\eq~(\ref{eq:appendix_EFT_Lagrangian}) becomes
\be
\label{eq:appendix_EFT_Lagrangian_canonical}
\mc L = -\f 1 2 (\p_\m a)^2 +a \f{f_I'}{f_I} 
\left( \p_\m a \p_\m h -\f a 2\f{f_I'}{f_I}(\p_\m h)^2 \right) 
+ \mc L_Q \ ,
\ee
where $'$ stands for differentiation wrt $h$, and 
\be
\label{eq:appendix_LQ}
\mc L_Q = \f{g_s^2a}{32\pi^2 f_I F_I}G^b_{\m\n}\widetilde G^{b\,\m\n} \ .
\ee

\subsection{Field-dependent noncanonical kinetic term}

Inspired by~\cite{Rigouzzo:2025hza}, we note that since the axion is of order
$H_I/2\pi$ during inflation, the time dependence of the kinetic term can be
neglected provided that
\be
\label{adiab}
\left\vert \f{f'_I}{f_I} \right\vert \ll H_I \ ,
\ee
which we will assume to be the case, and is what indeed happens in
Weyl-invariant Einstein-Cartan gravity discussed below (see also
\cite{Karananas:2025qsm}). In other words, the function should be varying
slowly over the Hubble times inflation lasts.\footnote{For instance, in Higgs
inflation, $f(h)=M_{\rm Pl}/\sqrt{1+\x_h h^2/M_{\rm Pl}^2}$ (see
below,~\eq~(\ref{eq:appendix_f})) meaning that $f_I \approx M_{\rm
Pl}^2/\sqrt{\x_h}h$, and $f_{\rm QCD}\approx M_{\rm Pl}$, so it changes by a
factor of 10 over the 60 inflationary efoldings. As a sanity check
$|f'_I/f_I| \approx h^{-1}\ll H_I$, as it should.} This guarantees that no
axion (over)production takes place at inflation, as the kinetic term for all
practical purposes can be taken to be canonical.

During reheating, the inequality~(\ref{adiab}) may not hold. In particular,
in Weyl-invariant Einstein-Cartan gravity    ${f'_I}/{f_I}\sim H_I$. We
checked, however, following~\cite{Garcia-Bellido:2012npk}, that for the
function $f_I$ increasing with time, as is the case in this theory, the axion
production is also negligible at reheating.

\subsection{Cutoff}

To find the cutoff, we should look at elastic tree-level scattering processes
due to the interaction of the axion with glue~(see the related discussion
in~\eg~\cite{Graf:2010tv, Salvio:2013iaa, DEramo:2021lgb}).

Focus on $\mc L_Q$ in~(\ref{eq:appendix_LQ}), and express the field strength
tensor in terms of gluons
\be
G^b_{\m\n} = \p_\m A^b_\n -\p_\n A^b_\m + g_s f^{bcd} A^c_\m A^d_\n \ ,
\ee
with $f^{bcd}$ the SU(3) structure constants; this yields
\be
\mc L_Q = \f{g_s^2 a}{16\pi^2 f_I F_I}\epsilon^{\m\n\rho\s}
\left(\p_\m A^b_\n \p_\rho A^b_\s 
+g_s f^{bcd}\p_\m A^b_\n A^c_\rho A^d_\s \right) \ .
\ee

From the above we notice that the process with the most problematic
high-energy behavior is $gg~\to~gg$ via the exchange of $a$
\begin{figure}[H]
	\centering
	\includegraphics[scale=.3]{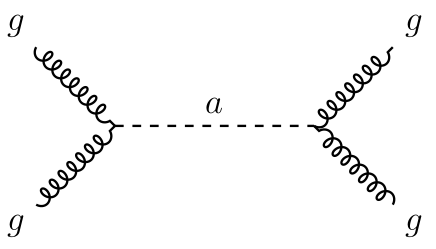}
\end{figure}
\noindent since each vertex contributes a factor of
\be
\f{g_s^2}{8\pi^2 f_I F_I} E^2 \ .
\ee

Requiring partial-wave unitarity of the corresponding amplitude
\be
\mc M \sim \f{g_s^4}{(8\pi^2 f_I F_I)^2}E^2 < 16\pi \ ,
\ee
we obtain
\be
\Lambda_I \sim \f{8\pi^{3/2}}{\a_s} f_I F_I 
\sim \mc O({\rm few})\times 10^3 f_I F_I \ ,
\ee
where we used  $g_s^2 = 4\pi \a_s$ and took $\a_s \sim 1/40$.

As a rule of thumb, we could have gotten a rough estimate of where to expect
the breakdown of the effective theory by plugging $a
\sim H_I/2\pi$ into~(\ref{eq:appendix_LQ}), which yields
\be
\mc L_Q \sim \f{\a_sH_I}{32\pi^2 f_I F_I} G^b_{\m\n}\widetilde G^{b\,\m\n} \ ,
\ee
meaning that
\be
\label{eq:appendix_naive_Lambda}
H_I < \f{32\pi^2f_I F_I}{\a_s} \sim 10^4 f_I F_I \ .
\ee
Evidently, $\Lambda_I$ is lower than, but nevertheless rather close to, the
naive estimate~(\ref{eq:appendix_naive_Lambda}).

\section{Onset of axion oscillations}
\label{app:onset}

In the main text we have tacitly assumed that the QCD-induced axion mass is
always the dominant one. Indeed, for most of the allowed values of the
explicit mass, this is justified. Given, however, the temperature dependence
of the QCD-induced axion mass---see below,
eq.~(\ref{eq:QCD_mass_temper})---it is in principle conceivable that at some
epoch it may become subdominant with respect to the explicit contribution.
For completeness we discuss how the picture is modified in this case.

Let us first recall what happens when no explicit mass term for the axion is
present. As the Universe expands, the Hubble scale eventually becomes
comparable to the QCD-induced axion mass, which triggers the onset of the
field oscillations; this happens when
\be
\label{eq:QCD_mass_Hubble_equal}
M_a(T_{\rm osc}) \simeq 3 H(T_{\rm osc}) \ .
\ee
The temperature-dependent axion mass can be approximated
as~\cite{OHare:2024nmr}~(see also~\cite{Marsh:2015xka})
\be
\label{eq:QCD_mass_temper}
M_a(T) = \Bigg\{\begin{array}{lr}
	M_a \left(\f{T_{\rm QCD}}{T}\right)^4 &~{\rm for}~T>T_{\rm QCD} \ ,\\
	M_a &~{\rm for}~T<T _{\rm QCD} \ , 
\end{array} 
\ee
where $T_{\rm QCD} \approx 150~{\rm MeV}$, and
\be
\label{eq:QCD_mass_fa}
M_a \simeq 5.7 \times 10^{-15}~{\rm GeV} 
\left(\f{10^{12}~{\rm GeV}}{f_a}\right) \ ,
\ee
has already appeared in the main text, see~\eq~(12). The Hubble parameter
during radiation domination reads
\be
\label{eq:Friedmann_eq}
H( T_{\rm osc}) = \f 1 3 \sqrt{\f{\pi^2}{10}g_\ast(T_{\rm osc})}
\f{T^2_{\rm osc}}{M_{\rm Pl}} \ ,
\ee
with $g_\ast$ the number of relativistic degrees of freedom at temperature
$T_{\rm osc}$. From~(\ref{eq:QCD_mass_Hubble_equal})-(\ref{eq:Friedmann_eq}),
and using that $g_\ast(T_{\rm osc})\approx 60$, we conclude that
\be
T_{\rm osc} \simeq \left(\f{10^{12}~{\rm GeV}}{f_a}\right)^{1/6}~{\rm GeV} \ ,
\ee
meaning that for admissible values of the decay constant, \ie $10^9~{\rm
	GeV}~\lesssim f_a \lesssim \times 10^{12}~{\rm GeV}$
\be
\label{eq:temp_osc_QCD}
T_{\rm osc} \sim \mc O({\rm few})~{\rm GeV} \ .
\ee

Let us now consider the effect of the explicit mass term $m_a$. Denote by
$\bar T_{\rm osc}$ the temperature at which the explicit mass alone would
trigger oscillations, \ie
\be
\label{eq:baremass_Hubble_equality}
m_a \simeq 3 H(\bar T_{\rm osc}) \ . 
\ee
Plugging~(\ref{eq:Friedmann_eq}) into~(\ref{eq:baremass_Hubble_equality}) and
solving for temperature, we find
\be
\label{eq:bare_Tosc}
\bar T_{\rm osc} \simeq g_\ast^{-1/4}(\bar T_{\rm osc})\sqrt{m_a M_{\rm Pl}}
\ .
\ee

Remember that $m_a$ is bounded both from above and below by requiring that:

(1)~strong CP violation does not exceed the experimentally measured value
\be
m_a \lesssim 10^{-5} M_a \ ;
\ee

(2)~domain walls disappear before BBN
\be
m_a \gtrsim 4.2\times 10^{-6} M_a 
\left(\f{f_a}{10^{12}~{\rm GeV}}\right)^{1/2} \ .
\ee
Using~(\ref{eq:QCD_mass_fa}), we find
\be
2.4 \times 10^{-20}~{\rm GeV}
\left(\f{10^{12}~{\rm GeV}}{f_a}\right)^{1/2}
\lesssim m_a \lesssim 
5.7 \times 10^{-20}~{\rm GeV}
\left(\f{10^{12}~{\rm GeV}}{f_a}\right)
\ee
In turn, from the above and~(\ref{eq:bare_Tosc}), it follows that 
\be
\label{eq:bare_Tosc_bounds}
0.24 g_\ast^{-1/4}(\bar T_{\rm osc}) 
\left(\f{10^{12}~{\rm GeV}}{f_a}\right)^{1/4}~{\rm GeV}
\lesssim \bar T_{\rm osc} \lesssim
0.37 g_\ast^{-1/4}(\bar T_{\rm osc}) 
\left(\f{10^{12}~{\rm GeV}}{f_a}\right)^{1/2}~{\rm GeV} \ .
\ee

It is reasonable to take the number of relativistic degrees of freedom to be
$\sim \mc O(10-60)$, meaning that at most $g_\ast^{-1/4}(\bar T_{\rm
	osc})\simeq 0.5$, thus
\be
0.1 \left(\f{10^{12}~{\rm GeV}}{f_a}\right)^{1/4}~{\rm GeV} 
\lesssim \bar T_{\rm osc} \lesssim 
0.2 \left(\f{10^{12}~{\rm GeV}}{f_a}\right)^{1/2}~{\rm GeV} \ .
\ee

For $f_a \simeq 10^{12}~{\rm GeV}$, we see that 
\be
\bar T_{\rm osc} \sim \mc O(10^{-1})~{\rm GeV}\ ,
\ee
what is below \eq~(\ref{eq:temp_osc_QCD}), while for $f_a \simeq 10^{9}~{\rm
	GeV}$, we find
\be
0.6~{\rm GeV} \lesssim \bar T_{\rm osc} \lesssim 6~{\rm GeV} \ ,
\ee
which are also below or of same order of magnitude as the temperature $T_{\rm
     osc}$~(\ref{eq:temp_osc_QCD}) set by the QCD mass. Therefore, for most
     of the allowed parameter space $M_a(\bar T_{\rm osc}) > m_a$, so the
     onset of oscillations is still controlled by the QCD contribution.

It is in the somewhat extreme situation on the boundary of the allowed
parameter space with $f_a$ saturating its lowest bound and $m_a$ its upper
bound, that the explicit mass becomes comparable to or larger than $M_a(\bar
T_{\rm osc})$. Even then, the temperature is increased only by a factor of
order one. This in turn (slightly) reduces the number of different domain
walls by at most a factor of few, since for $\bar T_{\rm osc} > T_{\rm osc}$
\be
\bar N \sim N \left(\f{T_{\rm osc}}{\bar T_{\rm osc}}\right)^{3/2} \ ,
\ee
with $N$ given in eq.~(16) in the main text. 
	
\section{Non-compact axion from Einstein-Cartan gravity}
\label{app:EinsteinCartan}

\subsection{Einstein-Cartan basics}

In the Einstein-Cartan framework, the gravitational interaction follows from
localizing the Poincar\'e group. The gauge fields associated with Lorentz
transformations and translations are the (spin) connection\,\footnote{Latin
letters stand for Lorentz indexes, which are manipulated with the Minkowski
metric $\eta_{AB} = {\rm diag}(-1,+1,+1,+1)$. } $\omega^{AB}_\m$ and tetrad
$e^A_\m$, respectively. Their corresponding field strength tensors are
curvature
\be
F^{AB}_{\m\n} = \p_\m \omega^{AB}_\n - \p_\n \omega^{AB}_\m 
+\omega^A_{\m C}\omega^{CB}_\n - \omega^A_{\n C}\omega^{CB}_\m \ ,
\ee
and torsion 
\be
T^A_{\m\n} = \p_\m e^A_\n - \p_\n e^A_\m 
+ \omega^A_{\m B}e^B_\n -\omega^A_{\n B}e^B_\m \ .
\ee

For our purposes it is useful to decompose torsion into its three irreducible
pieces
\be
v_\m = T^\n_{~\m\n} \ ,
~~~a^\m = E^{\m\n\rho\s}T_{\n\rho\s} \ ,
~~~\tau_{\m\n\rho} = \f 2 3 T_{\m\n\rho} 
-\f 1 3(v_\n g_{\m\rho}-v_\rho g_{\m\n}) 
-\f 1 3 (T_{\n\rho\m}-T_{\rho\n\m}) \ ,
\ee
with $T_{\m\n\rho}=e_{\m A}T^A_{\n\rho}$,~$E^{\m\n\rho\s}=\epsilon^{\m\n\rho\s}/{\rm det}(e^A_\m)$,~$g_{\m\n}=e^A_\m e_{\n A}$. 

In addition to the scalar curvature
\be
F = \f 1 4 e_{ABCD}E^{\m\n\rho\s}F^{AB}_{\m\n}e^C_\rho e^D_\s \ ,
\ee
there also exists a pseudoscalar curvature
\be
\widetilde F = E^{\m\n\rho\s}F^{AB}_{\m\n}e_{\rho A} e_{\s B} \ .
\ee

It is well-known that the above can be decomposed as
\bea
\label{eq:F_decomp}
F &=& R  +2 \nabla_\m v^\m - \f 2 3 v_\m^2 + \f{1}{24} a_\m^2 
+ \f 1 2 \tau_{\m\n\rho}^2 \ ,\\
\label{eq:Ftilde_decomp}
\widetilde F &=& - \nabla_\m a^\m + \f 2 3 a_\m v^\m 
- \f 1 2 E^{\m\n\rho\s}\tau_{\lambda\m\n}\tau^\lambda_{~\rho\s} \ ,
\eea
with $\nabla$ the torsion-free covariant derivative.

Under Weyl transformations, the gauge fields behave as
\be
\omega^{AB}_\m \mapsto \omega^{AB}_\m \ ,~~~e^A_\m \mapsto q^{-1}e^A_\m \ ,
~~~q=q(x) \ ,
\ee
meaning that
\be
F \mapsto q^{2}F \ ,~~~\widetilde F \mapsto q^2 \widetilde F \ , 
\ee
and
\be
v_\m \mapsto v_\m + 3q^{-1}\p_\m q \ ,
~~~a_\m \mapsto a_\m \ ,
~~~\tau_{\m\n\rho}\mapsto q^{-2}\tau_{\m\n\rho} \ .
\ee

Contrary to all other geometrical data, the torsion vector $v_\m$ transforms
inhomogeneously, and this is the reason why it can play the role of the Weyl
gauge field.

\subsection{Weyl-invariant Einstein-Cartan gravity}

Our starting point is the simplest possible instance of the Weyl-invariant
Einstein-Cartan gravity we constructed in~\cite{Karananas:2024xja}. The
Lagrangian reads
\be
\label{eq:WIEC_starting}
\mathcal L_{\rm EC} = \f{1}{g^2}F^2 +\f{1}{\widetilde g^2}\widetilde F^2 
-\f 1 2 \left(D_\m^W h\right)^2 - \f{\lambda  h^4}{4} \ ,
\ee
where $g$ and $\widetilde g$ are the gauge couplings of the Lorentz group,
$h$ the Higgs field in unitary gauge, $\lambda$ its self-coupling, and
\be
D_\m^W h = \p_\m h + \f 1 3 v_\m h \ ,
\ee
the Weyl-covariant derivative. Although obvious from the above, let us stress
that if the Lorentz gauge couplings vanish, there emerge accidental gauge
redundancies~\cite{Karananas:2024hoh,*Karananas:2024qrz} and the theory
becomes infinitely strongly coupled. As the latter sets the mass for the
axion, see later, in this setup the non-QCD tilt is a matter of
selfconsistency of the theory.

Importantly, note that in~(\ref{eq:WIEC_starting}) we could have included the
cross-term $F\widetilde F$, as well as a plethora of Weyl-invariant couplings
between the Higgs and gravitational
invariants~\cite{Karananas:2024xja,Karananas:2025ews,Karananas:2025fas,Karananas:2025qsm}.
We refrained from doing that, because we are interested in giving a minimal
working example. Although there is no difficulty in accounting for these
terms,
see~\cite{Karananas:2024xja,Karananas:2025ews,Karananas:2025fas,Karananas:2025qsm},
this will inevitably introduce unnecessary algebraic complications and
long(er) expressions. At the end of the day, the effect of these extra terms
in the limits of interest can be more-or-less captured by an overall
rescaling of the kinetic function $f$---given below
in~(\ref{eq:appendix_f})---by a constant.

How to express the above theory into its metric-equivalent form has been
shown in the aforementioned papers; we shall briefly outline the steps and
refer the interested reader there for details.
\begin{enumerate}
	\item~Write the Lagrangian~(\ref{eq:WIEC_starting}) in a first order form
	as
	\be
	\mc L_{\rm EC} = \chi^2 F + M_{\rm Pl}^2\phi \widetilde F -\f{g^2\chi^4}{4} 
	-\f{\widetilde g^2 M_{\rm Pl}^4 \phi^2}{4} -\f 1 2 \left(D_\m^W h\right)^2 
	- \f{\lambda h^4}{4}  \ ,
	\ee
	where $\chi$ and $\phi$ are Lagrange multipliers. For later convenience we took
	the former to have mass-dimension one and the latter to be dimensionless.
	
	\item~Fix the gauge $\chi=M_{\rm Pl}/\sqrt{2}$ and
	use~(\ref{eq:F_decomp},\ref{eq:Ftilde_decomp}), to obtain
	\bea
	\mathcal L_{\rm EC} &=& \f{M_{\rm Pl}^2}{2}R -\f 1 2 (\p_\m h)^2 - \f{\lambda h^4}{4} 
	- \f{g^2 M_{\rm Pl}^4}{16} - \f{\widetilde g^2 M_{\rm Pl}^4\phi^2}{4} 
	- \f 1 6 v^\m \p_\m h^2 - \f{1}{18}(6M_{\rm Pl}^2+h^2)v_\m^2 \nonumber \\
	&+&M_{\rm Pl}^2\left[ \nabla_\m v^\m +\f{1}{48}a_\m^2 
	+\f 1 4 \tau_{\m\n\rho}^2 -\phi\left(\nabla_\m a^\m -\f 2 3 a_\m v^\m 
	+\f 1 2 E^{\m\n\rho\s}\tau_{\lambda\m\n}\tau^\lambda_{~\rho\s}\right)\right] 
	\ .
	\eea
	
	\item~Find the equations of motion for $v,a,\tau$ and use them to integrate
	out torsion; this yields
	\bea
	\label{eq:nondiagonal}
	\mathcal L_{\rm EC} = \f{M_{\rm Pl}^2}{2}R &-& \f{\lambda h^4}{4} 
	- \f{g^2 M_{\rm Pl}^4}{16} 
	- \f{\widetilde g^2 M_{\rm Pl}^4\phi^2}{4} 
	-\f{3M_{\rm Pl}^2}{6M_{\rm Pl}^2(1+16\phi^2)+h^2}\times \nonumber \\
	&\times& \Big[(1+16\phi^2)(\p_\m h)^2 -16 h\phi \p_\m h \p^\m \phi 
	+4(6M_{\rm Pl}^2+h^2)(\p_\m \phi)^2 \Big] \ .
	\eea
	
	\item~Introduce
	\be
	\varphi = \f 1 2 
	\log\left[ \f{M_{\rm Pl}^2 \phi}{6M_{\rm Pl}^2+h^2}\right] \ ,
	\ee
	such that the kinetic sector of~({\ref{eq:nondiagonal}}) is diagonalized
	\bea
	\label{eq:diagonalized}
	\mathcal L_{\rm EC} &=& \f{M_{\rm Pl}^2}{2}R 
	-\f{3M_{\rm Pl}^2}{6M_{\rm Pl}^2+h^2}(\p_\m h)^2  -\f{\lambda h^4}{4} 
	\nonumber \\
	&&-\f{48(6M_{\rm Pl}^2+h^2)^2e^{4\varphi}}
	{M_{\rm Pl}^2+96(6M_{\rm Pl}^2+h^2)e^{4\varphi}}(\p_\m\varphi)^2
	- \f{\widetilde g^2}{4}(6M_{\rm Pl}^2+h^2)^2 e^{4\varphi} 
	- \f{g^2 M_{\rm Pl}^4}{16} \ .
	\eea
	
\end{enumerate}

Now, let us assume that $h\ll \sqrt{6} M_{\rm Pl}$~\cite{Karananas:2025qsm};
from~(\ref{eq:diagonalized}) we see that
\be
\mathcal L_{\rm EC} \approx \f{M_{\rm Pl}^2}{2}R - \f 1 2 (\p_\m h)^2 
-\f{\lambda h^4}{4} -\f{1728 M_{\rm Pl}^2
	e^{4\varphi}}{1+576e^{4\varphi}}(\p_\m\varphi)^2 
-9\widetilde g^2 M_{\rm Pl}^4 e^{4\varphi} - \f{g^2 M_{\rm Pl}^4}{16} \ ,
\ee
which in terms of 
\be
\theta = \sqrt{\f 3 2} {\rm arctanh}
\left[ \f{24e^{2\varphi}}{\sqrt{1+576 e^{4\varphi}}}\right] \ ,
\ee
becomes
\be
\mathcal L_{\rm EC} \approx \f{M_{\rm Pl}^2}{2}R - \f 1 2 (\p_\m h)^2 
-\f{\lambda h^4}{4} - \f{M_{\rm Pl}^2}{2}(\p_\m\theta)^2
-\f{\widetilde g^2 M_{\rm Pl}^4}{64}
{\rm sinh}^2\left(\sqrt{\f 2 3 \theta}\right) - \f{g^2 M_{\rm Pl}^4}{16}\ .
\ee
In turn, for $\theta \ll 1$, the above boils down to 
\be
\mathcal L_{\rm EC} \approx \f{M_{\rm Pl}^2}{2}R - \f 1 2 (\p_\m h)^2 
-\f{\lambda h^4}{4} - \f{M_{\rm Pl}^2}{2}(\p_\m\theta)^2
-\f{\widetilde g^2 M_{\rm Pl}^4}{96}\theta^2 - \f{g^2 M_{\rm Pl}^4}{16}\ .
\ee

As argued in~\cite{Karananas:2025qsm}, the selfconsistency of the theory at
the quantum level requires that the Higgs interact with the scalar curvature,
\ie the above be extended as
\be
\mathcal L_{\rm EC} =
\f{M_{\rm Pl}^2 + \x_h h^2}{2}R - \f 1 2 (\p_\m h)^2 
-\f{\lambda h^4}{4} - \f{M_{\rm Pl}^2}{2}(\p_\m\theta)^2
-\f{\widetilde g^2 M_{\rm Pl}^4}{96}\theta^2 - \f{g^2 M_{\rm Pl}^4}{16} \ ,
\ee
with $\x_h$ the so-called nonminimal coupling. In what follows we shall take
$\x_h \sim \mc O(10^4)$.

Moreover, requiring that Weyl symmetry be preserved at the quantum level,
selects the ``scale-invariant'' regularization
scheme~\cite{Englert:1976ep,Shaposhnikov:2008xi}, that subtracts divergences
in a manner that respects the classical symmetry. This is usually implemented
in Dimensional Regularization by continuing the theory to $D\neq 4$ spacetime
dimensions by multiplying all terms in the Lagrangian by functions of the
scalar fields. These give rise to evanescent operators, that in turn generate
vertices proportional to $D-4$; at the same time, the divergent pieces of the
loop integrals $\propto \f{1}{D-4}$, resulting into the appearance of
higher-dimensional operators that survive when $D\mapsto
4$~\cite{Shaposhnikov:2009nk}. Among those, there are terms that couple the
fields to the QCD topological invariant $Q$. Importantly, their exact form
cannot be fixed by first principles, and this is an ambiguity tied to the
(unknown) UV completion of the
theory~\cite{Karananas:2025ews,Shaposhnikov:2025znm}.

Let us, as a proof of concept, consider the simplest (but obviously not
unique) operator that we could think of
\be
\label{eq:appendix_higher_dimensional}
\mc O = \f{\theta}{32\pi^2 F} G^b_{\m\n}\widetilde G^{b\,\m\n} \ ,
\ee
where $F =~{\rm constant}$. 

As well-known, it is desirable to eliminate the nonminimal coupling and make
gravity canonical by means of a Weyl rescaling of the metric. The part of the
Lagrangian $\mc L_{\rm EC} + \mc O$ that contains $\theta$ only, becomes
\be
\mathcal L_{\rm EC} \supset -\f{f^2(h)}{2}(\p_\m\theta)^2 
- \f{f_m^2(h)\widetilde m_a^2}{2}\theta^2 
+ \f{\theta}{32\pi^2 F} G^b_{\m\n}\widetilde G^{b\,\m\n}  \ ,
\ee
where now
\bea
\label{eq:appendix_f}
&&f(h) = \f{M_{\rm Pl}}{\sqrt{1+\f{\x_h h^2}{M_{\rm Pl}^2}}} \ , \\ 
\label{eq:appendix_fm}
&&f_m(h) = \f{M_{\rm Pl}}{1+\f{\x_h h^2}{M_{\rm Pl}^2}} \ , \\ 
&&\widetilde m_a = \f{\widetilde g M_{\rm Pl}}{4\sqrt 3} \ .
\eea

From~(\ref{eq:appendix_f},\ref{eq:appendix_fm}), we can find the asymptotics
of these functions for inflation ($h\gg M_{\rm Pl}/\sqrt{\x_h}$)
\be
\label{eq:appendix_infl_limits}
f_I \approx \f{M_{\rm Pl}^2}{\sqrt{\x_h}h} \ ,~~~
f_{m,I} \approx \f{M_{\rm Pl}^3}{\x_h h^2} \ ,
\ee
and at low energies ($h\ll M_{\rm Pl}/\sqrt{\x_h}$) 
\be
\label{eq:appendix_QCD_limits}
~~~ f_{\rm QCD} \approx M_{\rm Pl} \ ,~~~
f_{m,{\rm QCD}} = f_{\rm QCD} \ .
\ee
It is clear that 
\be
f_I < f_{\rm QCD} \ ,~~~
f_{m,I} < f_{m,{\rm QCD}} \ .
\ee
As expected (we already mentioned that in the main text), both $f$ and $f_m$
are smaller during inflation due to the effect of the non-minimal coupling
that heavily suppresses these terms.

We now explicitly show that all requirements for the successful
implementation of our mechanism are satisfied in the above setup.

First of all, from~(\ref{eq:appendix_QCD_limits}) and the definition of the axionic decay constant $f_a = f_{\rm QCD} F$, we find
\be
\label{eq:appendix_F_fa}
F = \f{f_a}{M_{\rm Pl}} \ .
\ee

Second, the validity of the effective theory dictates that
\be
f_I F > 10^{-3} H_I \ ,
\ee
and using~(\ref{eq:appendix_F_fa}), yields
\be
\label{eq:appendix_fa_lower}
f_a > 10^{-3}H_I \f{\sqrt{\x_h}h}{M_{\rm Pl}} \ .
\ee

Third, to suppress isocurvature perturbations we saw that
\be
\f{H_I}{f_I} \gtrsim 2\sqrt 2 \pi F_{\rm QCD} \ ,
\ee
which upon plugging-in~(\ref{eq:appendix_F_fa}), we obtain
\be
\label{eq:appendix_fa_upper}
f_a \lesssim 10^{-1}H_I \f{\sqrt{\x_h}h}{M_{\rm Pl}} \ .
\ee

For metric Higgs inflation~\cite{Bezrukov:2007ep}
\be
\label{eq:appendix_Higgs_N}
\f{\sqrt{\x_h}h}{M_{\rm Pl}} \simeq \sqrt{\f{4N}{3}} \ ,
\ee
where $N\sim \mc O(50-60)$ is the number of efoldings, and the Hubble
parameter is~\cite{Bezrukov:2010jz}
\be
\label{eq:appendix_Hubble}
H_I \simeq \f{\sqrt\lambda M_{\rm Pl}}{\x_h} \simeq 10^{13}~{\rm GeV} \ .
\ee

Combining~(\ref{eq:appendix_fa_lower},\ref{eq:appendix_fa_upper}) and using~(\ref{eq:appendix_Higgs_N},\ref{eq:appendix_Hubble}) results into
\be
\label{eq:appendix_fa_combined}
\sqrt{\f{4N}{3}} \times 10^{12}~{\rm GeV} 
\gtrsim f_a \gtrsim 
\sqrt{\f{4N}{3}} \times 10^{10}~{\rm GeV} \ ,
\ee
that for $N=55$ gives
\be
8.6 \times 10^{12}~{\rm GeV} 
\gtrsim f_a \gtrsim 
8.6 \times 10^{10}~{\rm GeV} \ ,
\ee
which is fully consistent with the phenomenologically interesting window.

Finally, the bounds on strong CP violation translate into bounds for the
Lorentz gauge coupling $\widetilde g$,
\be
6.8\times 10^{-38} \left(\f{10^{12}~{\rm GeV}}{f_a}\right)^{1/2}
\lesssim \widetilde g \lesssim 
1.6\times 10^{-37} \left(\f{10^{12}~{\rm GeV}}{f_a}\right) \ , 
\ee
and provided that the axion comprises the entirety of DM, then $f_a \simeq
3.2\times 10^{11}~{\rm GeV}$ and the above gives
\be
\widetilde g \sim \mc O(10^{-37}) \ .
\ee

\end{document}